\documentclass[11pt,twoside]{book} 
\usepackage{asp2008n}
\usepackage{times}
\usepackage{lscape}
\usepackage{epsf}

\usepackage{graphicx}
\usepackage{amssymb}
\usepackage{longtable}
\usepackage[figuresright]{rotating}

\newcommand{\ts}{\thinspace}
\newcommand{\about}    {$\sim$\ts}
\newcommand{\aboutless}{$\simless$\ts}

\def\etal {{ et al.~}}

\newcommand{\simless}{\mathbin{\lower 3pt\hbox {$\rlap{\raise 5pt\hbox{$\char'074$}}\mathchar"7218$}}}

\newcommand{\IRAS}{{\it IRAS}}

\mainmatter

\newlength{\deftabcolsep}
\begin{document}

\title{The Monoceros R2 Molecular Cloud}
\author{John M. Carpenter}
\affil{California Institute of Technology, Pasadena, CA 91125, USA}
\author{Klaus W. Hodapp}
\affil{Institute for Astronomy, University of Hawaii, Hilo, HI 96720, USA}

\begin{abstract} The Monoceros R2 region was first recognized as a chain of
reflection nebulae illuminated by A- and B-type stars. These nebulae are
associated with a giant molecular cloud that is one of the closest massive star
forming regions to the Sun. This chapter reviews the properties of the Mon~R2
region, including the namesake reflection nebulae, the large scale molecular
cloud, global star formation activity, and properties of prominent star forming
regions in the cloud.

\end{abstract}

\section{The Mon R2 Reflection Nebulae}

The Monoceros R2 region is distinguished by a chain of reflection
nebulae that extend over 2\deg\ on the sky. The brightest of these
nebulae were studied as far back as \citet{Seares20}, and were
included in a larger study of nebulous objects by \citet{Hubble22},
who demonstrated that the extended emission can be attributed to the
associated stars. The nebulae are included in the Catalog of Bright
Diffuse Galactic Nebulae constructed by \citet{Cederblad46}.
Various lists of nebulae identified from inspection of the Palomar
Observatory Sky Survey appear in \citet{DG63,DG66}, \citet{VDB66}, and
\citet{Herbst76}. \citet{GGD78} identified seven additional
nebulous objects as candidate Herbig-Haro objects, although they were later
shown to be reflection nebulae \citep{Cohen80}. The nomenclature
``Mon~R2'' originated with \citet{VDB66} to indicate the second association
of reflection nebulae in the constellation
Monoceros\footnote{The Mon~R1 reflection nebulae are
part of the Mon~OB1 association, which includes the Galactic cluster
NGC~2264.}.

\begin{table}
\caption{Optical Nebulae in the Mon~R2 Region\label{tbl:nebula}}
\smallskip
\begin{center}
{\small
\begin{tabular}{lcclrrrr}
\tableline
\noalign{\smallskip}
Source &
$\alpha$ &
$\delta$ &
SpT$^a$  &
vdB      &
DG       &
Ced      &
NGC      \\
\cline{2-3} \\
& \multicolumn{2}{c}{(J2000)}\\
\noalign{\smallskip}
\tableline
\noalign{\smallskip}
HR  1     & 6:06:58 & $-5$:55:10 &         &    &    &    &      \\
HR  2     & 6:07:26 & $-6$:38:13 & A0:     &    &    &    &      \\
HR  3     & 6:07:30 & $-6$:29:16 &         &    &    &    &      \\
HR  4     & 6:07:32 & $-6$:24:03 & B2 V    & 67 & 88 & 63 & 2170 \\
HR  5a    & 6:07:46 & $-6$:21:47 & A5:     &    &    &    &      \\
HR  5b    & 6:07:46 & $-6$:23:02 &         &    &    &    &      \\
HR  6     & 6:07:49 & $-6$:16:35 &         &    &    &    &      \\
HR  7     & 6:07:48 & $-6$:26:46 &         &    &    &    &      \\
HR  8$^b$ & 6:08:05 & $-6$:21:34 & B1 V    & 69 & 90 & 66 &      \\
HR  9$^b$ & 6:08:04 & $-6$:13:38 & B2 V    & 68 & 89 & 65 &      \\
HR 10     & 6:08:26 & $-5$:20:19 & B1 V    & 70 & 91 &    &      \\
HR 11     & 6:09:15 & $-6$:30:29 &         &    &    &    &      \\
HR 12     & 6:09:25 & $-6$:18:35 &         &    &    &    &      \\
HR 13     & 6:09:31 & $-6$:19:36 & B3 V    & 72 & 93 & 68 & 2182 \\
HR 14     & 6:09:45 & $-6$:18:35 & B5 V    &    &    &    &      \\
HR 15     & 6:10:01 & $-6$:18:42 & B8 V    &    &    &    &      \\
HR 16     & 6:10:15 & $-6$:27:32 &         &    &    &    &      \\
HR 17     & 6:10:37 & $-6$:09:58 &         &    &    &    &      \\
HR 18     & 6:10:47 & $-6$:12:41 &         &    & 94 & 69 & 2183 \\
HR 19     & 6:10:56 & $-6$:14:22 &         &    &    &    &      \\
HR 20     & 6:10:57 & $-6$:16:21 &         &    &    &    &      \\
HR 21     & 6:11:00 & $-6$:14:36 &         &    &    &    &      \\
HR 22     & 6:11:06 & $-6$:12:33 & B8-A0   &    & 95 & 70 & 2185 \\
HR 23     & 6:11:49 & $-6$:09:22 & B4 V    & 74 & 96 & 71 &      \\
HR 24     & 6:12:45 & $-6$:12:31 &         &    &    &    &      \\
HR 25     & 6:12:46 & $-6$:10:49 &         &    &    &    &      \\
HR 26     & 6:14:46 & $-6$:20:36 &         &    &    &    &      \\
HR 27     & 6:14:53 & $-6$:22:44 & B7 V    &    & 97 &    &      \\
HR 28     & 6:15:21 & $-6$:25:50 & B5 V    &    & 98 &    &      \\
HR 29$^c$ & 6:12:50 & $-6$:13:11 & K4      &    &    &    &      \\
HR 30     & 6:10:58 & $-6$:14:39 & Be      &    &    &    &      \\
DG 92     & 6:08:30 & $-6$:30:30 &         &    & 92 &    &      \\
GGD 11    & 6:08:33 & $-6$:18:16 &         &    &    &    &      \\
GGD 12    & 6:10:45 & $-6$:12:45 &         &    &    &    &      \\
GGD 13    & 6:10:46 & $-6$:13:04 &         &    &    &    &      \\
GGD 14    & 6:10:50 & $-6$:12:01 &         &    &    &    &      \\
GGD 15    & 6:10:50 & $-6$:11:15 &         &    &    &    &      \\
GGD 16    & 6:12:44 & $-6$:15:24 &         &    &    &    &      \\
GGD 17    & 6:12:50 & $-6$:12:57 &         &    &    &    &      \\
\noalign{\smallskip}
\tableline
\noalign{\smallskip}

\multicolumn{8}{l}{\parbox{0.8\textwidth}{\footnotesize $^a$
   Spectral types from \citet{Herbst76} and \citet{Carballo92}.
}}\\
\multicolumn{8}{l}{\parbox{0.8\textwidth}{\footnotesize $^b$
   \citet{Herbst76} associated HR~8 with vdB~69 and HR~9 with vdB~68.
   These designations are inconsistent with the coordinates of these
   sources and the finding chart in that paper (cf. Figure~8 in
   Downes \etal~1975). We have maintained the same designations in
   \citet{Herbst76}, and have updated the finding chart in
   Fig.~\ref{fig:poss} accordingly.
}}\\
\multicolumn{8}{l}{\parbox{0.8\textwidth}{\footnotesize $^c$
   T~Tauri star also known as Bretz~4 \citep{her72}.
}}\\
\multicolumn{8}{l}{\parbox{0.8\textwidth}{\footnotesize
   Ced : \citet{Cederblad46}\\
   DG  : \citet{DG63}\\
   GGD : \citet{GGD78}\\
   HR  : \citet{Herbst76}\\
   vdB : \citet{VDB66}\\
}}\\

\end{tabular}
}
\end{center}
\end{table}

\begin{figure}
\begin{center}
\includegraphics[angle=0,height=0.9\textheight]{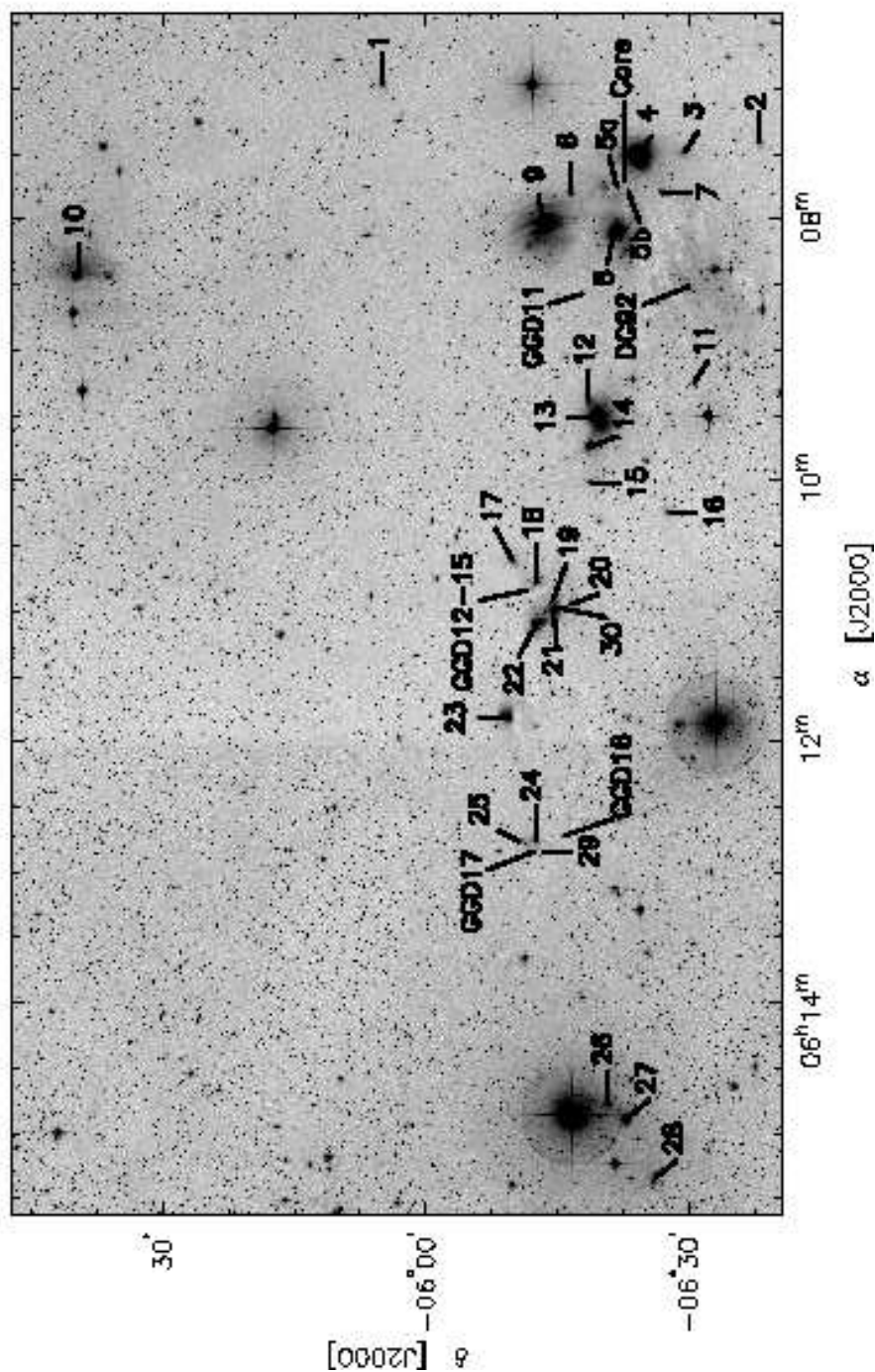}
\end{center}
\vspace{-5mm}
\caption{
   Location of the optical nebulae in Mon~R2 (see Table~\ref{tbl:nebula})
   marked on the blue print from the Palomar Observatory Sky Survey.
   The regions numbered 1-30 refer to the sources cataloged by
   \citet{Herbst76}. The ``GGD'' and ``DG'' sources are the nebulous
   objects identified by \citet{GGD78} and \citet{DG63} respectively.
   The main Mon~R2 star forming cloud core is also indicated.
  \label{fig:poss}
}
\end{figure}

\begin{figure}[tb]
\begin{center}
\includegraphics[angle=-90,scale=0.54]{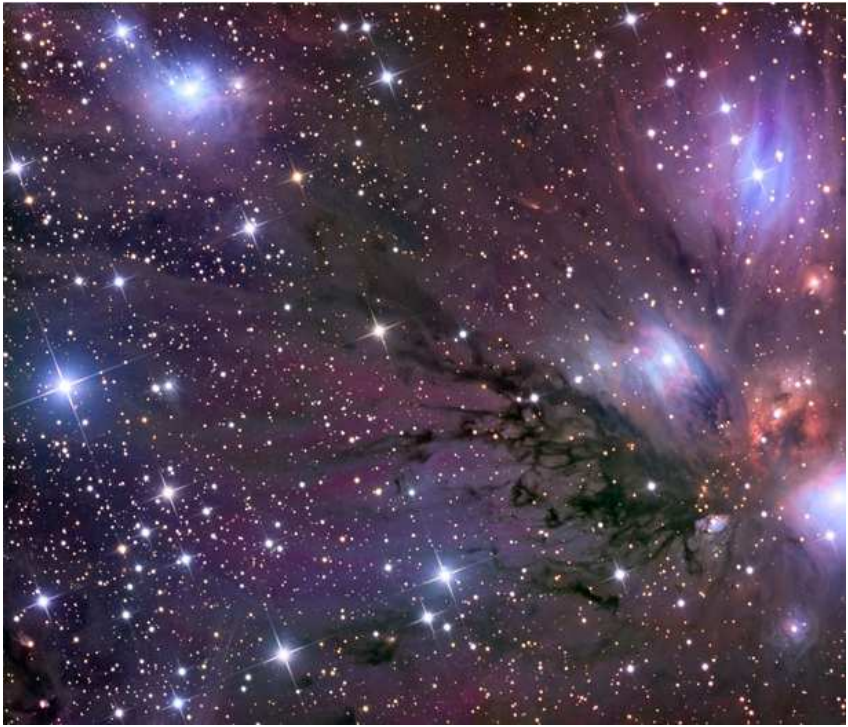}
\end{center}
\caption{
   Optical photograph of the Mon~R2 region that highlights the presence of
   reflection nebulae, embedded star forming regions, and dark nebulae.
   The Mon~R2 core is the red region on the right between the NGC~2170
   (vdB~67; lower right) and vdB~69 (center right) blue reflection nebulae.
   North is to the upper right, and east is to the upper left.
   Photograph courtesy of R. Croman.
  \label{fig:croman}
}
\end{figure}

Table~\ref{tbl:nebula} lists the nebulous sources in the Mon~R2 region
identified in the studies mentioned above. Coordinates were determined from
the digitized Palomar Sky Survey using previously published finding charts
\citep{Downes75,Herbst76,Carballo88} and coordinate lists
\citep{Rodriguez80,Carballo92}. Figure~\ref{fig:poss} marks the location
of the nebulae on the blue print from the Palomar Observatory Sky Survey. The
sources GGD 12 through 15 are located within a small area of the sky and
are often collectively referred to as GGD~12-15. The most extensively
studied region is the active star forming site embedded in a dense molecular
core (identified as the ``core'' in Fig.~\ref{fig:poss}) that lies midway
between the reflection nebulae vdB~67 and vdB~69 (see Fig.~\ref{fig:croman} for
a more detailed optical color image of this region).
In many contexts, the name ``Mon~R2'' refers to this particular star
forming region.

\citet[see also Herbst \& Racine~1976]{Racine68} conducted the first detailed
spectroscopic and photometric study of the Mon~R2 nebulae and found that the
illuminating stars have mainly B spectral types, where B1~V is the earliest
type star in the association. \citet{Herbst76} estimated
an age of 6 to 10~Myr for the association. The lower limit is ascertained
since the color-magnitude diagram for the Mon~R2 stellar population does not
deviate from the zero-age main-sequence for stars with $(B-V)_0 \le 0.05$~mag,
while a pre-main-sequence population is observed for such stars in other
clusters such as the Orion Nebula Cluster and NGC~2264. The upper age limit was
inferred since the B1 stars in the association still lie on the main-sequence.
As noted by \citet{Herbst76}, the young age for the association suggests that
the B-stars likely formed within the cloud, and the reflection nebulae are not
the result of a chance encounter of B-stars with a gas-cloud (c.f. the
Pleiades; White~2003).

\citet{Racine68} estimated a distance of 830$\pm$50~pc to Mon~R2 using
spectroscopy and/or $UBV$ photometry for 10 stars in the reflection nebulae.
Using similar techniques, \citet[][as reported in Downes \etal~1975]{RK68}
derived a distance of 700~pc. \citet{Racine70} revised the distance
from \citet{Racine68} to 950~pc, but did not provide further details on the
factors leading to the new estimate. \citet{Herbst76} fitted the
zero-age-main-sequence locus from \citet{Johnson63} to dereddened $UBV$
photometry and also obtained a distance of 830$\pm$50~pc, which is the
distance typically adopted for the Mon~R2 association.

\section{The Mon R2 Molecular Cloud}
\label{cloud}

\begin{figure}[htb]
\begin{center}
\includegraphics[angle=-90,scale=0.50]{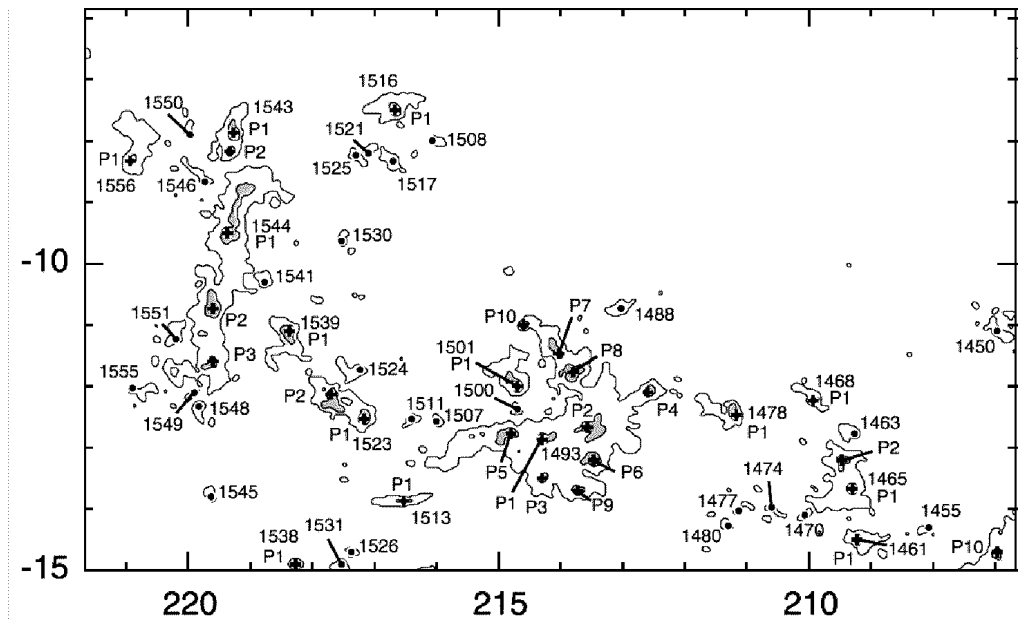}
\end{center}
\caption{
   Extinction map in Galactic coordinates of the Mon~R2 cloud and vicinity as
   derived from star counts in the Digital Sky Survey~I database
   \citep{Dobashi05}. The lowest contour indicates the $A_V = 0.5$~mag
   boundary. Symbols indicate identified clouds (filled circles) and clumps
   within clouds (plus signs). The Mon~R2 cloud is object TGU~1493 in the
   \citet{Dobashi05} catalog at ($\ell, b \approx 214.3^\circ, -12.9^\circ$),
   and includes \citet{Lynds62} dark nebulae L1643, 1644, 1645, and 1646. The
   main star forming core is near clump P2, and the GGD~12-15 region is
   near P8.
  \label{fig:dobashi}
}
\end{figure}

In addition to bright reflection nebulae, the Mon~R2 region is readily
identified on optical photographs as an opaque region in silhouette against
the backdrop of field stars (see Fig~\ref{fig:croman}). \citet{Lynds62}
identified 4
prominent ``dark nebulae'' (L1643,
1644,
1645, and
1646) that form the main
part of the molecular cloud that is associated with the reflection nebulae.
An extinction map of the Mon~R2 region derived from optical star counts by
\citet{Dobashi05} is presented in Figure~\ref{fig:dobashi} that clearly
delineates the Mon~R2 region (labeled TGU~1493 in this map) and dark clouds in
its vicinity.

\subsection{Global Characteristics}

\begin{figure}[!tb]
\begin{center}
\includegraphics[angle=0,width=\textwidth]{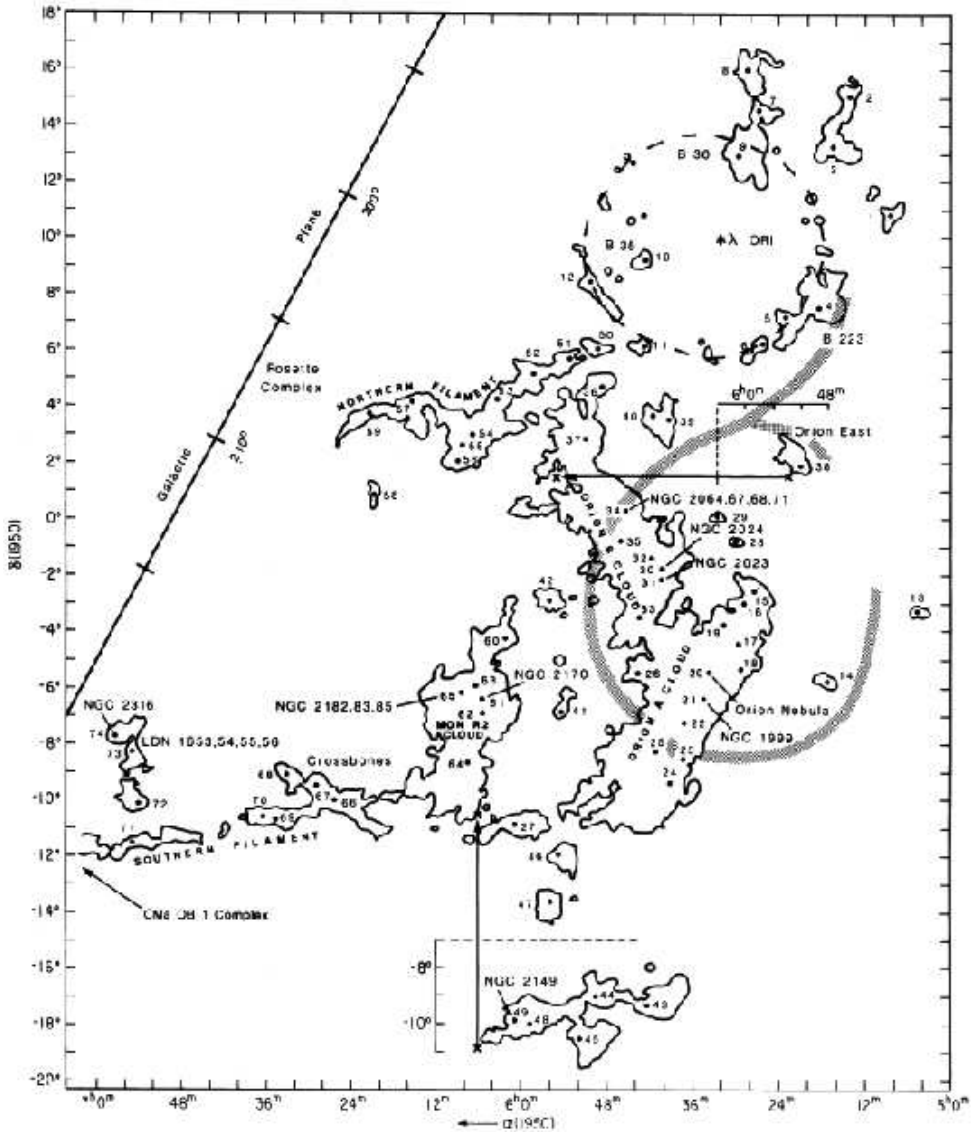}
\end{center}
\caption{
  Overview of the molecular clouds in the vicinity of Mon~R2 from
  \citet{Maddalena86}. The contour show the integrated CO line
  emission between $-10$ and 20~km~s$^{-1}$ at a level of 1.28~K~km~s$^{-1}$
  The dots with numbers correspond to CO emission peaks
  listed in Table~1 in \citet{Maddalena86}.
  \label{fig:comad}
}
\end{figure}

\begin{figure}[p]
\begin{center}
\includegraphics[height=0.9\textheight]{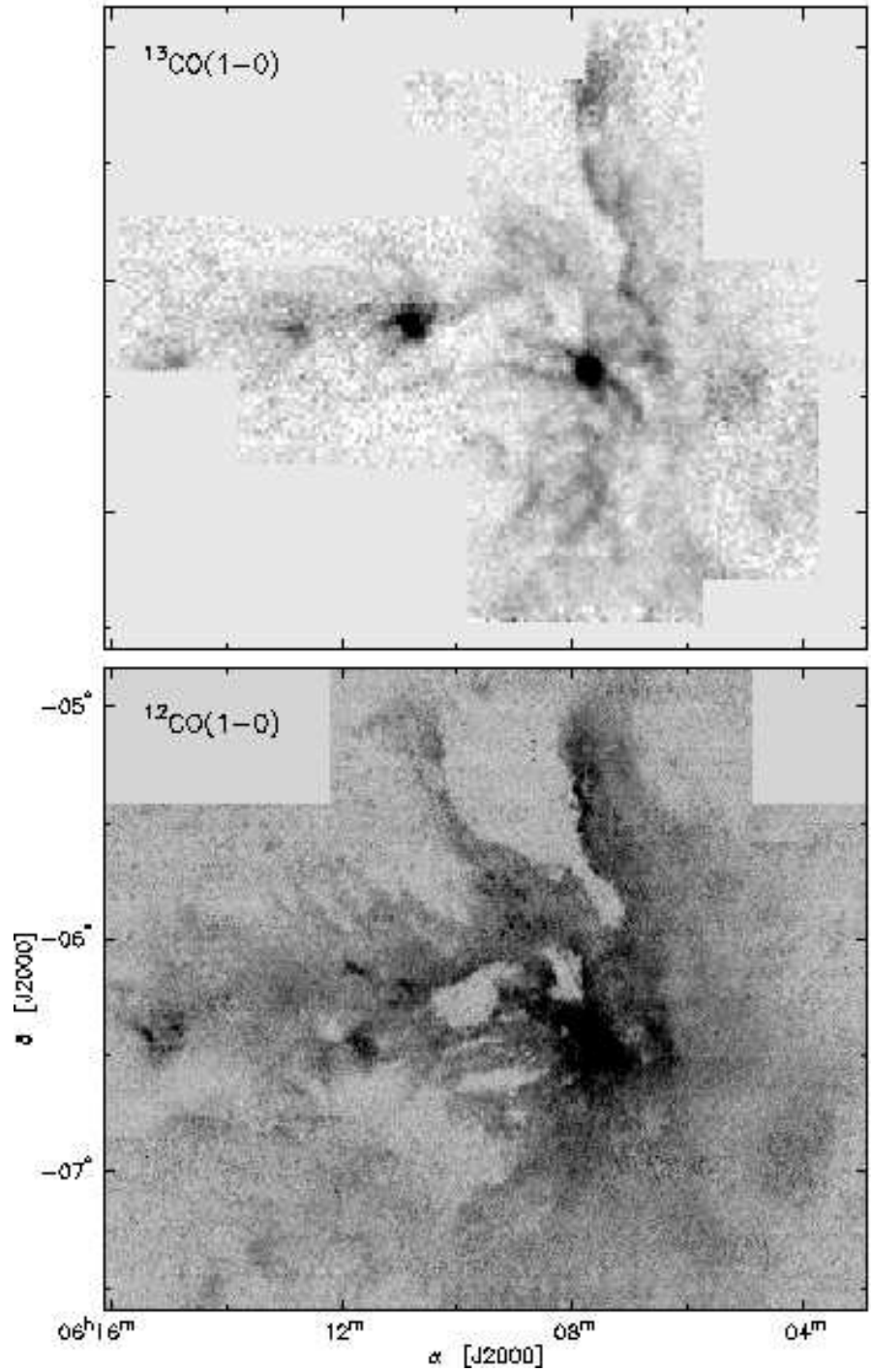}
\end{center}
\caption{
   Molecular line maps of the peak $^{12}$CO J=1-0 antenna temperature
   \citep[bottom panel;][]{Xie92} and the $^{13}$CO J=1-0 integrated intensity
   \citep[top panel;][]{Miesch99}. Darker gray scales corresponds to
   regions of brighter emission.
  \label{fig:co}
}
\end{figure}

The Mon~R2 region received renewed interest with the first observations of
carbon monoxide in the interstellar medium and the discovery of molecular
clouds \citep{Wilson70}. \citet{Loren74} reported the first $^{12}$CO (J=1-0)
detection in Mon~R2, and \citet{Kutner75} showed that at least five of the
reflection nebulae are associated with local maxima in $^{12}$CO maps.
A number of maps have subsequently been made, with ever improving
resolution and coverage, and increasing number of molecular species
\citep{Loren77,Maddalena86,Montalban90,Xie94,Miesch99,Kim04,Wilson05}.
A large-scale view of the molecular clouds in Orion and Monoceros as traced
by carbon monoxide is shown in Figure~\ref{fig:comad} \citep{Maddalena86}.
The Mon~R2 cloud extends over a $\sim 3^\circ\times6^\circ$
($44~{\rm pc}\times88~{\rm pc}$) region roughly parallel to the Galactic plane,
but offset from the plane by \about --12\deg. Mon~R2 is located in projection
near the Orion system of molecular clouds, although Orion is a factor of two
closer in distance. \citet{Heiles98} has suggested that the two regions may
nonetheless be linked by an expanding superbubble (GSH 238+00+09)
that triggered star formation in these regions.

Figure~\ref{fig:co} presents images of the J=1--0 emission from $^{12}$CO
\citep{Xie94} and $^{13}$CO \citep{Bally91,Miesch99}. The $^{12}$CO map,
tracing approximately the kinetic temperature, shows a bright peak near the
Mon~R2 core and a sharp filament of gas extending in nearly a north-south
direction. The $^{13}$CO image traces the molecular column density and
shows several peaks in the molecular emission that correspond to the location
of reflection nebulae. \citet{Xie94} analyzed the global velocity field in the
Mon~R2 cloud as traced by $^{12}$CO and suggested the cloud is an expanding
shell of molecular gas with a dynamical time scale of \about 4~Myr. No clear
energy source driving the expansion was identified, although the center of the
shell is near the NGC~2182 reflection nebula. \citet{Xie92} estimated a cloud
mass of $4\times 10^4$ M{$_{\odot}$} for the region shown in Figure~\ref{fig:co} that
has been observed in both $^{12}$CO and $^{13}$CO using traditional LTE
analysis \citep{Dickman78}. The maps produced by \citet{Maddalena86} encompass
the entire cloud, and they estimated a higher mass of $9\times 10^4$ M{$_{\odot}$} by
adopting a constant conversion factor from $^{12}$CO J=1-0 integrated
intensity to H$_2$ column density \citep[see, e.g.,][]{Strong88}.

\subsection{Stellar Content}

The global star formation activity in the Mon~R2 molecular cloud has been
explored at multiple wavelengths, including
(1) ROSAT X-ray pointed observations to search for young stars,
(2) 2MASS near-infrared and Spitzer mid-infrared observations to identify
    stellar clusters,
(3) \IRAS\ far-infrared images to probe for embedded star forming regions,
(4) radio continuum surveys to identify embedded massive stars,
and
(5) H$_2$ imaging surveys to trace jets originating from young stars.
The stellar content traced by these diagnostics paint a consistent picture
where the Mon~R2 and GGD~12-15 region are the two most active star forming
sites in the Mon~R2 molecular cloud. We now summarize the results from many
of these observations.

\begin{figure}[htbp]
\begin{center}
\includegraphics[angle=-90,scale=0.74]{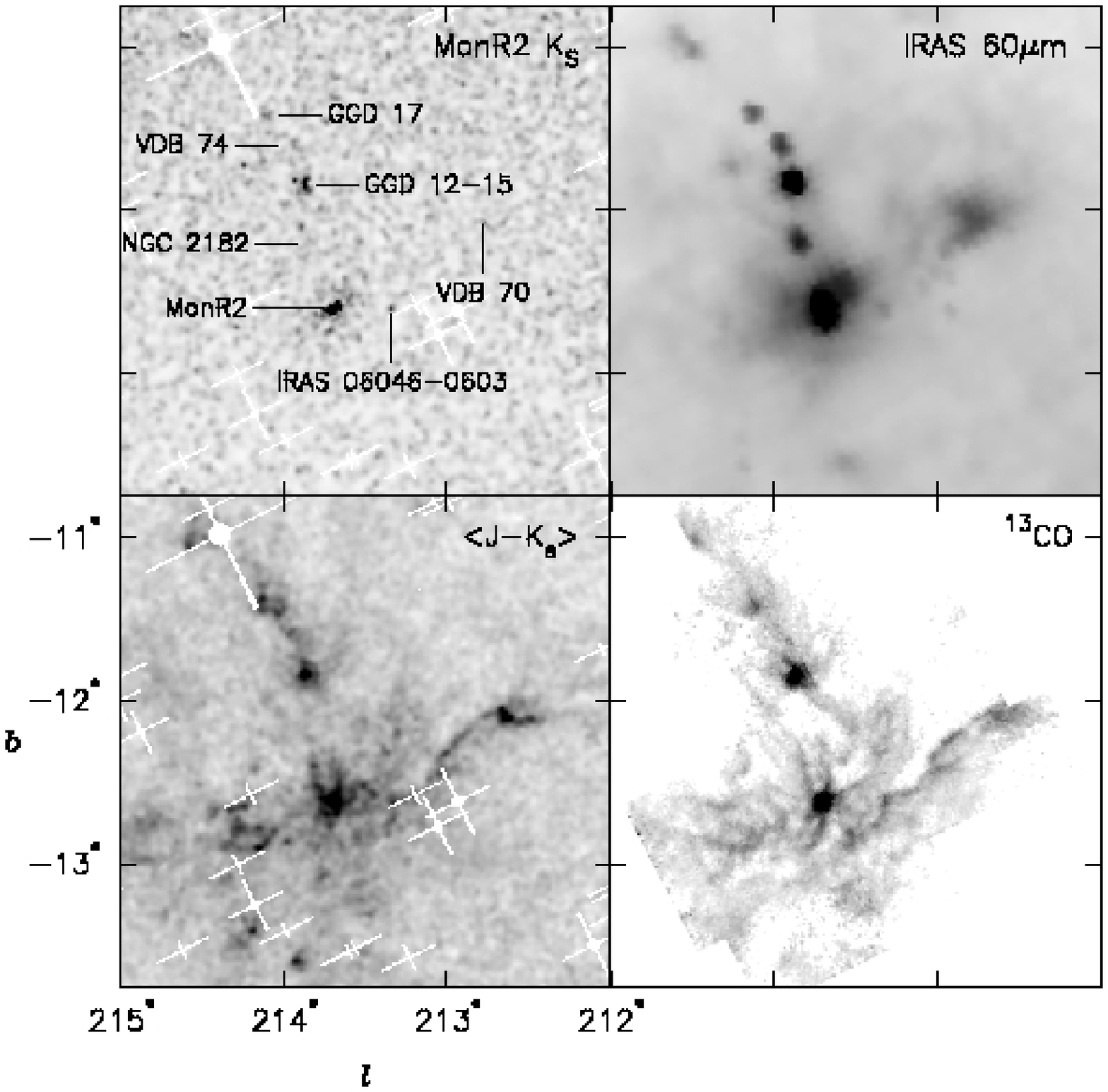}
\end{center}
\caption{
   {\it Upper left:}
    $K_{\rm s}$-band stellar surface density map of the Mon~R2
    molecular cloud for stars with magnitudes of 6.0~mag $\le m(K_{\rm s}) \le
    14.3$~mag.
  {\it Upper right:}
    The \IRAS\ 60~\micron\ image displayed in a logarithmic stretch.
  {\it Lower left:}
    An image of the average $J-K_{\rm s}$ color for stars observed by 2MASS.
  {\it Lower right:}
    A map of the integrated $^{13}$CO(J=1-0) intensity map \citep{Miesch99}.
  In each panel, darker halftones represent higher intensities. In the
  $K_{\rm s}$-band density map and the average $J-K_{\rm s}$ color image,
  the white ``crosses'' are regions around bright stars that were masked out
  in generating the 2MASS catalog. The sources labeled vdB are reflection
  nebulae cataloged by \citet{VDB66}, and sources labeled GGD are from the
  list of small nebulae noted by \citet{GGD78}. Figure from
  \citet{Carpenter00}.
  \label{fig:monr2_kernel}
}
\vspace{-4mm}
\end{figure}

The lower mass stellar content in the Mon~R2 cloud has been investigated by
\citet{Carpenter00}, who used the 2MASS point source catalog to identify compact
stellar clusters and quantify any stellar population distributed more
uniformly over the cloud. Four clusters were found based on enhancements
in the stellar surface density relative to the field star population as
shown in Figure~\ref{fig:monr2_kernel}.
These four clusters, listed in
Table~\ref{tbl:clusters}, are associated with the Mon~R2 core (containing 371
stars brighter than $K_{\rm s}$=14.3 after subtracting the expected field star
population), GGD~12-15 (134 stars), GGD~17 (23 stars), and IRAS~06046-0603
(15 stars). \citet{Carpenter00} also assessed the magnitude of any
distributed stellar population by measuring the surface density of stars toward the
molecular cloud relative to the field star surface density. The number of
stars contained in the distributed population is between 190 and 790 stars.
These results suggest that an appreciable portion of the stars in the Mon~R2
cloud are contained in just two clusters, but also that a substantial isolated stellar
population may exist.
\begin{table}[tb]
\caption{Embedded Stellar Clusters in the Mon~R2 Molecular Cloud\label{tbl:clusters}}
\smallskip
\begin{center}
\begin{tabular}{lcc@{\extracolsep{10pt}}c@{\extracolsep{10pt}}c@{\extracolsep{0pt}}rc}
\tableline
\noalign{\smallskip}
\multicolumn{1}{c}{Region} &
\multicolumn{2}{c}{Galactic} &
\multicolumn{2}{c}{Equatorial (J2000)} &
\multicolumn{1}{c}{N$_{\rm stars}$} &
R$_{\rm eff}$ \\
\cline{2-3}
\cline{4-5}
         &
$\ell$   &
$b$      &
$\alpha$ &
$\delta$ &
         &
[pc]\\
\noalign{\smallskip}
\tableline
\noalign{\smallskip}
IRAS 06046-0603 & 213.3381 & $-$12.6029  & 6:07:08.1 & $-$6:03:53 &  15 & 0.41\\
MonR2           & 213.6955 & $-$12.5926  & 6:07:47.8 & $-$6:22:20 & 371 & 1.85\\
GGD 12-15       & 213.8745 & $-$11.8410  & 6:10:49.1 & $-$6:11:38 & 134 & 1.13\\
GGD 17          & 214.1337 & $-$11.4173  & 6:12:48.0 & $-$6:13:56 &  23 & 0.61\\
\noalign{\smallskip}
\tableline
\end{tabular}
\end{center}
\end{table}

\begin{figure}[htb]
\begin{center}
\includegraphics[angle=-90,scale=1.0]{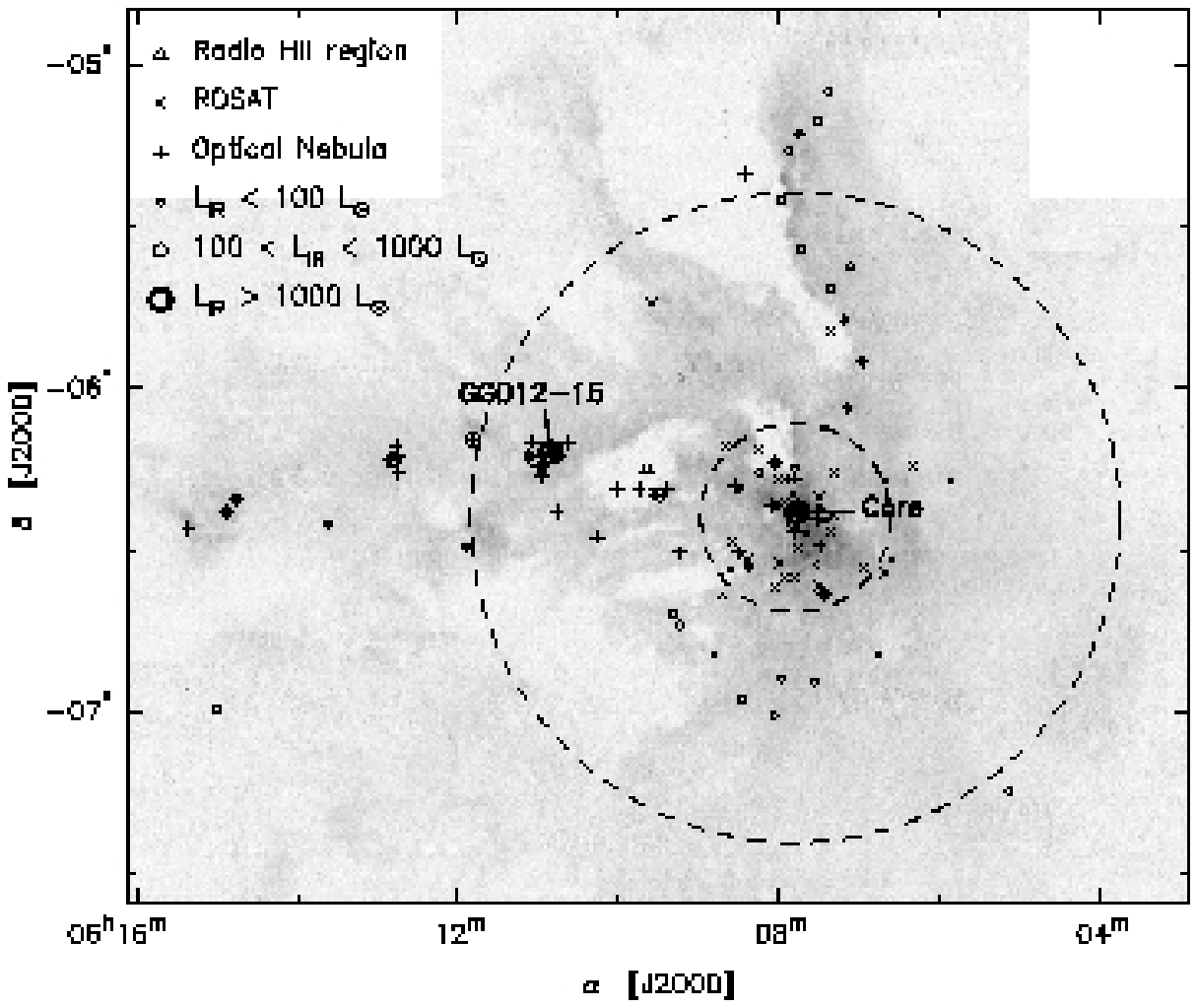}
\end{center}
\caption{
  Spatial distribution of various tracers of star formation activity overlaid
  on an image of the peak $^{12}$CO intensity \citep{Xie92}.
  The \ion{H}{II} regions are from \citet{Hughes85}, the ROSAT sources
  from \citet{G98}, and the optical nebulae from Table~\ref{tbl:nebula}.
  Open circles represent the \IRAS\ point sources listed in
  Table~\ref{tbl:iras}, where the size of the circle is proportional
  to the infrared luminosity ($L_{IR}$). The dashed circles show
  the full field of view of the ROSAT pointed observations (outer circle),
  and the 35$'$ diameter region that was imaged with high sensitivity and
  resolution (inner circle).
  \label{fig:co_iras}
}
\end{figure}

\citet{G98} used {\it ROSAT} to conduct an X-ray survey over a 2\deg\ diameter
region centered on the Mon~R2 cloud. The highest resolution and
sensitivity was achieved for the inner \about 35\arcmin. They detected
41 point sources with a signal to noise ratio greater than 2.5, and
possibly seven additional detections at lower confidence levels. The spatial
distribution of the 41 X-ray sources are indicated in Figure~\ref{fig:co_iras}.
Based on the observed X-ray hardness ratios and the ratio of the X-ray to
stellar bolometric luminosity, \citet{G98} suggest that most of the detected
sources are analogous to Herbig Ae/Be stars and the more X-ray active T~Tauri
stars. Extended X-ray emission was detected as well, which may originate from
partially resolved X-ray emitting young stars within the Mon~R2 cloud.

\IRAS\ provided a sensitive survey of the embedded stellar content over the
entire Mon~R2 molecular cloud. \citet{Xie92} selected a sample of 36 \IRAS\
point sources having photometric properties consistent with star forming
regions \citep[see][]{Beichman86} in that
(1) the sources are detected at 25~\micron\ or both 60~\micron\ and
    100~\micron,
and
(2) $S_\nu(25~\micron) \ge S_\nu(12~\micron)$.
Table~\ref{tbl:iras} lists the 36 \IRAS\ sources
meeting these criteria. The far-infrared luminosity for each source was
estimated as
$L_{IR}(L_\odot) = 4.7\times10^{-6}\:D^2\:(
      {S_\nu(12~\mu m)\over0.79} +
      {S_\nu(25~\mu m)\over2.0} +
      {S_\nu(60~\mu m)\over3.9} +
      {S_\nu(100~\mu m)\over9.9})$,
where $D$ is the distance in parsecs and $S_\nu$ is the flux density in
Janskys \citep{Casoli86,Parker91}.
Column~8 in Table~\ref{tbl:iras} lists the \IRAS\ luminosity computed from the
four \IRAS\ bands. For sources where an upper limit to the flux density is
listed in the \IRAS\ catalog for one or more bands, column~9 lists the upper
limit to the luminosity. The luminosities range from 1.5~L$_\odot$\ to
26,000~L$_\odot$. The two most luminous \IRAS\ sources are
associated with the Mon~R2 core ($L_{IR}$ \about 26,000~L$_\odot$) and GGD~12-15
(5700~L$_\odot$). Most of the remaining \IRAS\ sources have luminosities
less than 1000~L$_\odot$. Figure~\ref{fig:co_iras} shows the location of the
\IRAS\ sources overlaid on a $^{12}$CO image, where the diameter of
the circle is proportional to the far-infrared luminosity. The \IRAS\ sources
are spatially distributed in two groups. One group extends east-west along the
chain of reflection nebulae. The second group extends north-south along the
sharp boundary of the cloud traced by $^{12}$CO emission.

\setlength{\tabcolsep}{0.49\deftabcolsep}
\begin{table}[htb]
\begin{center}
\smallskip
\caption{IRAS Point Sources in the Mon~R2 Molecular Cloud$^*$\label{tbl:iras}}
{\footnotesize
\begin{tabular}{l@{\hskip8pt}r@{\hskip8pt}r@{\hskip1pt}r@{\hskip1pt}r@{\hskip1pt}r@{\hskip4pt}r@{\hskip8pt}r@{\hskip8pt}r}
\tableline
\noalign{\smallskip}
& & & \multicolumn{4}{c}{S$_\nu$} & &\\
IRAS PSC &
\multicolumn{1}{c}{$\alpha_{2000}$} &
\multicolumn{1}{c}{$\delta_{2000}$} &
\multicolumn{1}{c}{12\,$\mu$m} &
\multicolumn{1}{c}{25\,$\mu$m} &
\multicolumn{1}{c}{60\,$\mu$m} &
\multicolumn{1}{c}{100\,$\mu$m} &
\multicolumn{1}{c}{L$_{\rm IR}$} &
\multicolumn{1}{c}{L$_{{\rm IR}u}$} \\
\noalign{\smallskip}
\tableline
\noalign{\smallskip}
06027-0714 & 06:05:08.2 & -07:14:42  & $<$0.25 & 0.96 & 4.93 & 6.35 &  $>$7.7 & $<$8.7\\
06045-0554 & 06:06:58.5 & -05:55:08  & 0.76 & 0.99 & 0.98 & 4.96 & 7.1\\
06046-0536 & 06:07:06.7 & -05:37:24  & $<$0.25 & 0.25 & 3.42 & $<$9.49 &  $>$3.2 & $<$7.4\\
06046-0603 & 06:07:08.5 & -06:03:47  & $<$0.25 & 2.24 & 15.3 & 28.3 &  $>$25.6 & $<$26.6\\
06047-0546 & 06:07:11.1 & -05:47:21  & $<$0.25 & 1.66 & 4.83 & 3.96 &  $>$8.0 & $<$9.0\\
06049-0541 & 06:07:21.4 & -05:41:38  & 0.36 & 1.84 & 4.47 & $<$9.49 &  $>$8.2 & $<$11.3\\
06049-0504 & 06:07:23.4 & -05:04:54  & $<$0.29 & 0.36 & 1.16 & $<$10.3 &  $>$1.5 & $<$6.1\\
06050-0623 & 06:07:27.5 & -06:23:47  & 11.8 & 119 & $<$442 & $<$20190 &  $>$240 & $<$7210\\
06050-0509 & 06:07:31.6 & -05:10:21  & $<$0.39 & 1.48 & 4.18 & 10.3 &  $>$9.2 & $<$10.8\\
06051-0653 & 06:07:33.5 & -06:54:26  & $<$0.25 & 0.25 & 2.18 & $<$5.21 &  $>$2.2 & $<$5.0\\
06052-0533 & 06:07:43.5 & -05:34:17  & $<$0.25 & 0.60 & 13.1 & 22.3 &  $>$19.1 & $<$20.2\\
06052-0512 & 06:07:45.0 & -05:12:40  & $<$0.25 & 0.68 & 2.50 & $<$15.9 &  $>$3.2 & $<$9.4\\
06053-0622 & 06:07:46.7 & -06:23:00  & 470 & 4095 & 13070 & 20190 & 26008\\
06053-0614 & 06:07:48.9 & -06:14:44  & 0.78 & 1.07 & $<$6.05 & $<$81.8 &  $>$4.9 & $<$36.7\\
06054-0515 & 06:07:52.7 & -05:16:04  & 0.84 & 1.37 & $<$2.41 & $<$34.0 &  $>$5.7 & $<$18.8\\
06055-0524 & 06:07:58.9 & -05:25:03  & $<$0.45 & 0.44 & 4.80 & $<$10.4 &  $>$4.7 & $<$9.9\\
06055-0653 & 06:07:58.0 & -06:53:45  & 0.52 & 2.11 & 3.85 & 4.76 & 10.3\\
06056-0621 & 06:08:03.8 & -06:21:38  & 7.07 & 29.3 & $<$13070 & $<$20190 &  $>$76.4 & $<$17530\\
06056-0700 & 06:08:04.2 & -07:00:38  & $<$0.26 & 0.69 & 2.24 & $<$11.2 &  $>$3.0 & $<$7.7\\
06058-0615 & 06:08:14.8 & -06:15:33  & 0.71 & 0.83 & 18.4 & $<$155 &  $>$19.5 & $<$70.0\\
06059-0632 & 06:08:23.6 & -06:33:02  & 0.38 & 4.42 & 9.64 & 6.52 & 18.9\\
06060-0657 & 06:08:27.7 & -06:57:42  & 0.92 & 1.57 & $<$3.59 & $<$9.74 &  $>$6.3 & $<$12.5\\
06060-0617 & 06:08:29.5 & -06:18:26  & 0.28 & 1.23 & $<$18.4 & 39.0 &  $>$15.9 & $<$31.2\\
06068-0643 & 06:09:14.5 & -06:43:57  & 0.73 & 1.35 & 1.91 & $<$8.47 &  $>$6.8 & $<$9.5\\
06068-0641 & 06:09:19.8 & -06:41:55  & 0.98 & 2.00 & $<$1.91 & $<$8.47 &  $>$7.2 & $<$11.6\\
06070-0619 & 06:09:30.0 & -06:19:40  & 2.95 & 15.2 & 178 & 314 & 287\\
06084-0611 & 06:10:51.0 & -06:11:54  & 27.1 & 604 & 3613 & 4876 & 5682\\
06085-0613 & 06:10:57.8 & -06:14:37  & 2.99 & 3.27 & $<$3613 & $<$4876 &  $>$17.5 & $<$4612\\
06086-0611 & 06:11:07.5 & -06:12:32  & 2.38 & 2.57 & $<$3613 & $<$4876 &  $>$13.9 & $<$4608\\
06093-0608 & 06:11:48.6 & -06:09:30  & 6.60 & 12.3 & 141 & 249 & 245\\
06094-0628 & 06:11:53.3 & -06:29:20  & $<$0.25 & 0.32 & 7.03 & 25.2 &  $>$14.6 & $<$15.6\\
06103-0612 & 06:12:48.3 & -06:13:19  & 4.02 & 20.8 & 70.3 & 123 & 149\\
06111-0624 & 06:13:36.2 & -06:25:01  & 0.44 & 0.48 & 0.69 & $<$11.2 &  $>$3.1 & $<$6.8\\
06123-0619 & 06:14:44.9 & -06:20:24  & 0.36 & 0.42 & $<$1.87 & 21.2 &  $>$9.1 & $<$10.6\\
06124-0621 & 06:14:53.1 & -06:22:43  & 1.82 & 2.07 & 15.6 & 53.8 & 41.4\\
06125-0658 & 06:15:00.8 & -06:59:15  & $<$0.25 & 0.19 & $<$0.51 & 4.12 &  $>$1.6 & $<$3.1\\
\noalign{\smallskip}
\tableline
\noalign{\smallskip}
\multicolumn{9}{l}{\parbox{0.8\textwidth}{\footnotesize $^*$
   Flux densities are in units of Janskys and luminosities are in L$_\odot$.
}}\\[2ex]
\end{tabular}
}
\end{center}
\end{table}
\setlength{\tabcolsep}{\deftabcolsep}

\citet{Hughes85} traced the massive embedded stellar content in the
\linebreak Mon~R2
molecular cloud by surveying a 16~deg$^2$ region in the radio continuum
at 3.2~GHz and 10.55~GHz with angular resolution of $8\arcmin.3$ and
$2\arcmin.8$, respectively. The sensitivity limit of the survey was 40~mJy
at both frequencies, sufficient to detect an \ion{H}{II} region excited by a
B1 star. Three radio continuum sources with a thermal spectrum were detected.
The survey also detected 10 additional radio continuum sources with
a non-thermal spectrum that were presumed extragalactic in origin. An
additional ten sources were detected in only one frequency, and are most
likely extragalactic based on the expected radio continuum source counts.

Table~\ref{tbl:hii} lists the positions and flux densities of the three
\ion{H}{II} regions detected by \citet{Hughes85}. The location of the
\ion{H}{II} regions are shown as open triangles in Figure~\ref{fig:co_iras}.
One of the radio continuum sources is located in the Mon~R2 core, and a second
in the GGD~12-15 region. The third radio continuum source, which has not been
well studied, is situated in a hole in the $^{12}$CO emission
(see Fig.~\ref{fig:co_iras}) and has no clear correspondence with a reflection
nebula or \IRAS\ source. Assuming that the radio continuum emission is
optically thin and the ionization flux is produced by a single star at the
distance of Mon~R2, the spectral type of the ionizing star for all three radio
continuum sources is an early B-type star as listed in Table~\ref{tbl:hii}.

\begin{table}[htb]
\caption{Compact \ion{H}{II} Regions in the Mon~R2 Cloud from Hughes \& Baines (1985)\label{tbl:hii}}
\smallskip
\begin{center}
{\footnotesize
\begin{tabular}{rcc@{\extracolsep{3pt}}r@{\extracolsep{4pt}}r@{\extracolsep{4pt}}rr}
\tableline
\noalign{\smallskip}
\multicolumn{1}{r}{Source}  &
$\alpha$     &
$\delta$     &
\multicolumn{2}{c}{$S_\nu$ (mJy)}&
\multicolumn{1}{c}{SpT} &
\multicolumn{1}{c}{Region} \\
\cline{2-3}
\cline{4-5}
& \multicolumn{2}{c}{(J2000)} & (3.2 GHz) & (10.5 GHz) \\
\noalign{\smallskip}
\tableline
\noalign{\smallskip}
G213.7--12.6\tablenotemark{a} & 06:07:46.2 & $-6$:23:09.3 & 6200$\pm$200 & 6600$\pm$200 & B0 to O9.5 & Mon~R2 core \\
G213.8--12.1\tablenotemark{b} & 06:09:39   & $-6$:15.1    &  120$\pm$10  &  115$\pm$15  & B0.5 to B1 \\
G213.9--11.8\tablenotemark{c} & 06:10:50.6 & $-6$:11:49.8 &   95$\pm$20  &  100$\pm$15  & B0.5 to B1 & GGD 12-15 \\
\noalign{\smallskip}
\tableline
\noalign{\smallskip}
\multicolumn{7}{l}{\parbox{0.8\textwidth}{\footnotesize $^a$
    Coordinates from \citet{Wood89}} }\\
\multicolumn{7}{l}{\parbox{0.8\textwidth}{\footnotesize $^b$
    Coordinates from \citet{Hughes85} and uncertain by
  30$''$} }\\
\multicolumn{7}{l}{\parbox{0.8\textwidth}{\footnotesize $^c$
    Coordinates from \citet{Kurtz94}} }\\

\end{tabular}
}
\end{center}
\end{table}

\citet{Downes75} showed that the radio continuum emission in the Mon~R2 core
extends over a 2\arcmin\ diameter region, but most of the emission
originates from a compact \ion{H}{II} region. High resolution observations by
\citet[see also Wood \& Churchwell~1989]{Massi85} showed that this compact
component is nearly circular with a diameter of 27\arcsec\ and a bright arc
of radio emission present along one edge. \citet{Massi85} modeled the radio
emission as a blister-type \ion{H}{II} region where the ionizing star is on the
surface of the molecular cloud. The radio continuum emission in the GGD~12-15
region exhibits a cometary structure with a maximum extent of
\about 4-5\arcsec\ \citep{Kurtz94,Gomez98}. \citet{Gomez98} investigated the
kinematics of the ionized gas and suggested that the \ion{H}{II} region is
undergoing a champagne flow.

\subsection{Magnetic Field}

The pattern of polarization angles measured toward field stars was used by
\citet{Zartisky87}, \citet{Hodapp87}, and later by \citet{Yao97} to map the
projected magnetic field direction in the Mon~R2 molecular cloud.
\begin{figure}[htb]
\begin{center}
\includegraphics[angle=0,width=0.9\textwidth]{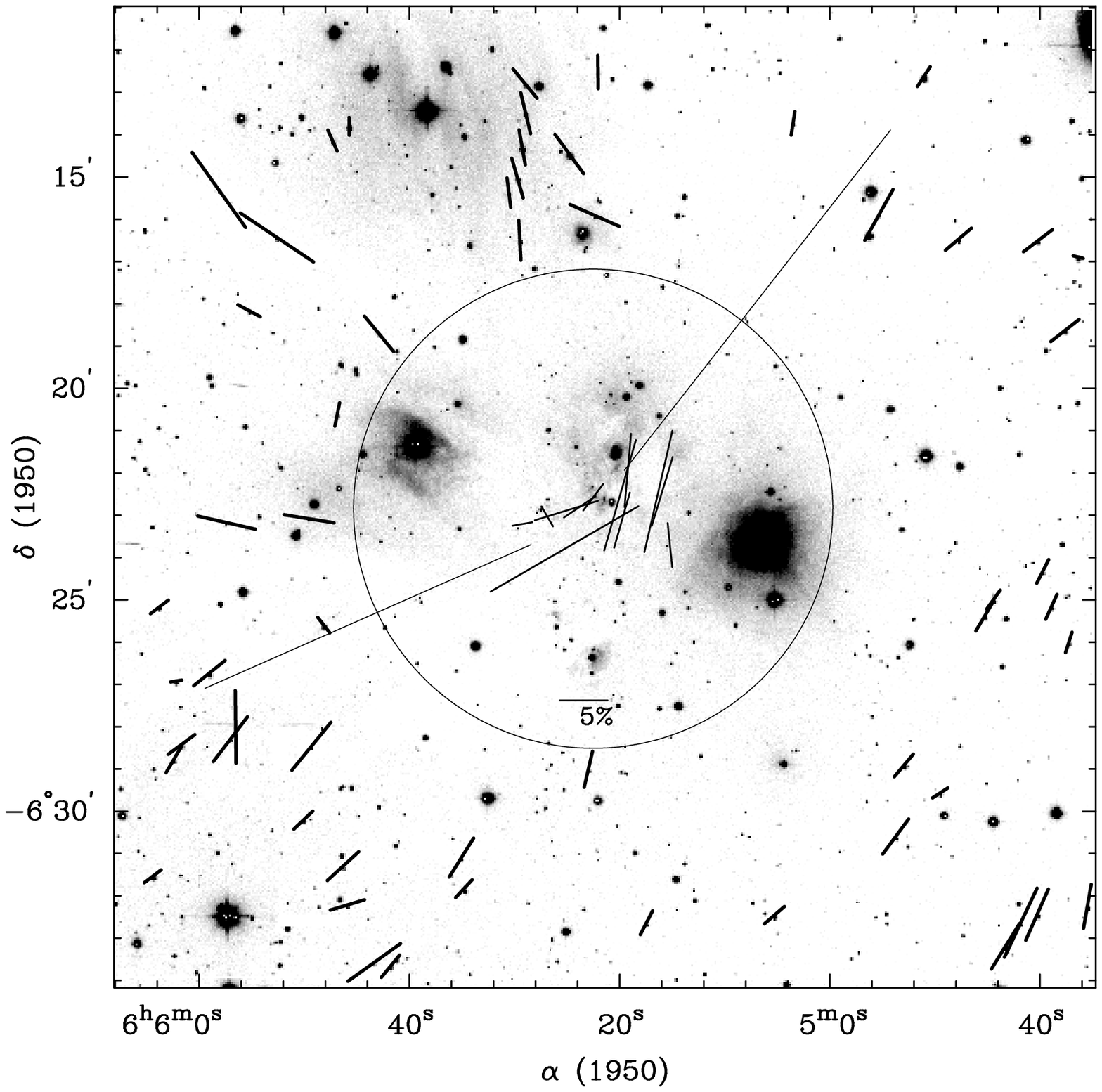}
\end{center}
\caption{
  $R$-band image of the Mon~R2 core region overlaid with $R$-band polarization
  vectors \citep{Jarrett94}.
  The circle represents the area where \citet{Jarrett94} did not obtain
  polarization data. In the inner 3$'$ the $I$- and $K$-band point source
  measurements from \citet{Hodapp87} were added. Note that
  Fig.~\ref{fig:monr2_hodapp} from \citet{Hodapp87} shows mostly the
  polarization vectors of scattered light in the reflection nebula; those
  polarization vectors are not included here, since they do not trace the
  magnetic field orientation. The long lines intersecting the circle represent
  the long-axis of the molecular outflow from the core region \citep{Bally83}.
  \label{fig:jarrett}
}
\end{figure}
The $R$-band measurements over a 9$'$ field by \citet{Zartisky87} show a
projected magnetic field orientation of P.A. = 167$^\circ$, parallel to the
local
Galactic magnetic field, and also parallel to the direction of the Mon~R2
molecular outflow \citep{Bally83}. The studies by \citet{Hodapp87} based on
$I$- and $K$-band polarization vectors, and by \citet{Yao97} based on
$K_{\rm s}$-band data, concentrated on the region near the cloud core. Both
suggested that the projected magnetic field appears to have an hourglass
shape, indicative of the collapse of a supercritical cloud where the frozen-in
magnetic field lines are bent toward the cloud's center by the gravitational
field of the cloud core. The conclusion that the magnetic field in Mon~R2 is
bent was confirmed by sub-millimeter (800~$\mu$m) emission polarimetry by
\citet{Greaves95} and far-infrared (100~$\mu$m) data from \citet{Novak89}.

A study of the polarization pattern in the $R$-band over a much wider field
using three 23$'\times23'$ CCD images by
\citet[][see Fig.~\ref{fig:jarrett}]{Jarrett94} images confirmed
many of the earlier results and demonstrated the continuity between the
magnetic field orientation measured in the tenuous parts of the cloud (by
$R$-band polarimetry) and the dense parts (measured by $I$-band and $K$-band
polarimetry). However, they also showed a second pattern of polarization
vectors east of the molecular core that cannot be explained simply by
the gravitational pull of the molecular core. Instead, their results are best
interpreted in a scenario where a large-scale expanding shell has distorted
the magnetic field lines extending from the core to the north.

\section{Individual Regions}

In this section we review the properties of individual star forming regions
in the Mon~R2 molecular cloud, including the main Mon~R2 core
(see Sect.~\ref{monr2}), the small group of nebulae in GGD~12-15
(Sect.~\ref{ggd12}), and the star forming sites associated with HH~866
(Sect.~\ref{HH866}) and GGD~16-17 (Sect.~\ref{ggd17}).

\begin{figure}[htb]
\begin{center}
  \includegraphics[angle=0,width=0.9\textwidth]{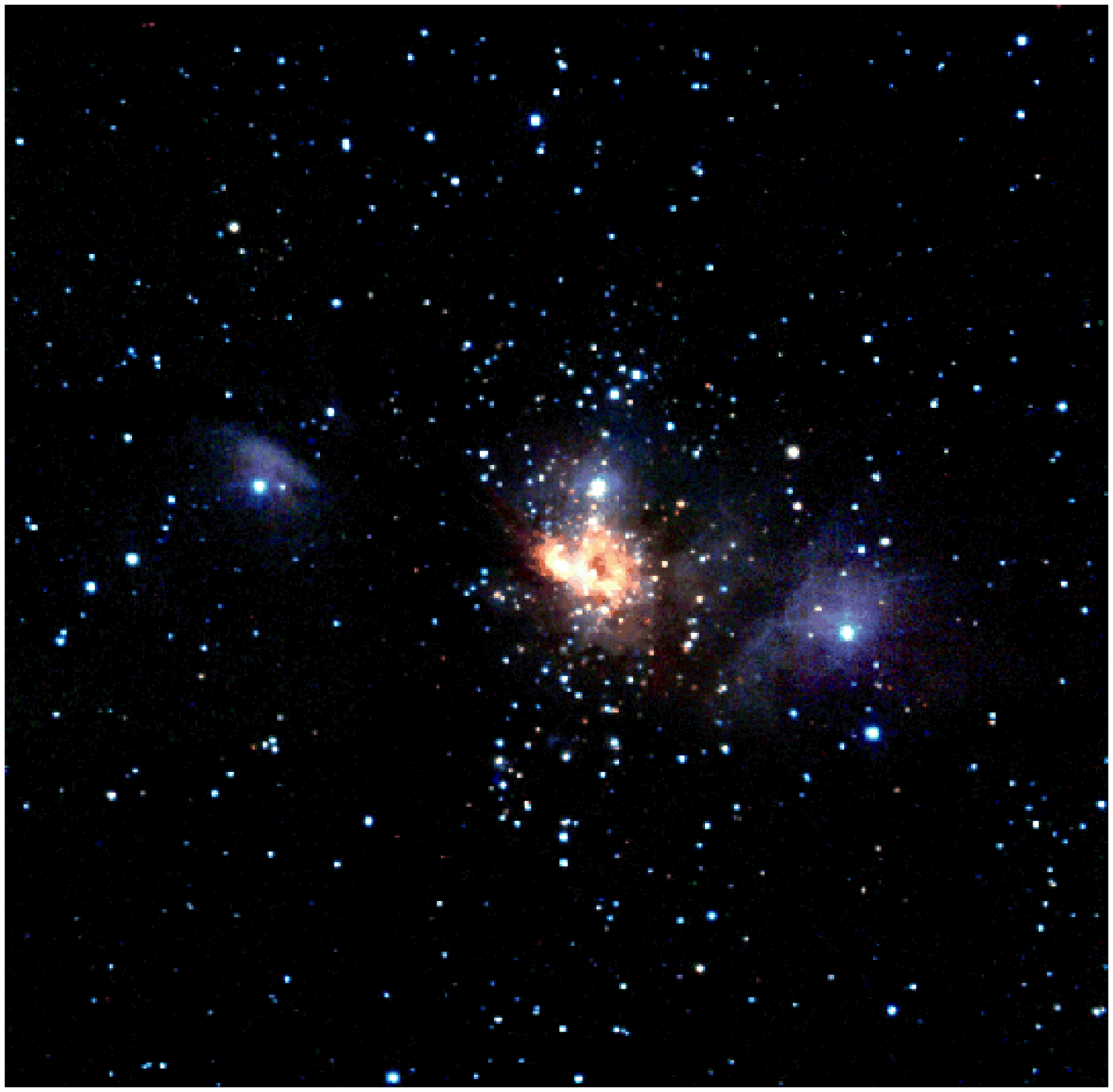}
\end{center}
\caption{
  Three-color composite (blue -- $J$; green -- $H$; red -- $K$) of the Mon~R2
  cluster \citep{Carpenter97}. The field of view of the image is
  \about $15'\times15'$; north is up and east to the left.
  The reflection nebulae vdB~69 and vdB~67 are located to the left and right,
  respectively, of the central Mon~R2 cluster. This infrared image can be
  compared with the optical image presented in Fig.~\ref{fig:croman}.
  \label{fig:monr2_color}
}
\end{figure}

\subsection{The Mon~R2 Cluster}
\label{monr2}

The Mon~R2 core is distinguished by a stellar cluster
(see Fig.~\ref{fig:monr2_color}), one of the largest
(6.6~pc) and most powerful known molecular outflows
\citep{Bally83,Wolf90,MRL91,Xie93,Tafalla94},
a compact \ion{H}{II} region
\citep{Downes75,Massi85}, H$_2$O and OH masers \citep{Downes75,Knapp76},
and X-ray emission \citep{Hamaguchi00,Kohno02,Nakajima03}.

Molecular line \citep{Tafalla97,Choi00} and submillimeter continuum
\citep{Walker90,Henning92,G97} observations have shown that the cluster is
embedded in a dense core which has a diameter of \about 3\arcmin\ (0.7~pc).
The core displays a rich chemical structure that is driven
by the high ultraviolet flux from the embedded B-star \citep{Rizzo05}. The
core mass derived from multi-transition CS molecular line observations is
\about 760 M{$_{\odot}$}, of which 130 M{$_{\odot}$} is being accelerated by an outflow
\citep[][where masses and sizes are scaled to a distance of
830~pc]{Tafalla97}. \citet{Walker90} and \citet{Henning92} estimate a core
mass of \about 200 M{$_{\odot}$} within the central 2.7~arcmin$^2$ based on
millimeter continuum observations.

\citet{Beckwith76} discovered the embedded cluster in the Mon~R2 core
through $H$, $K$, 10~\micron, and 20~\micron\ scan-maps.
\begin{figure}[htb]
\begin{center}
\includegraphics[angle=-90,scale=0.51]{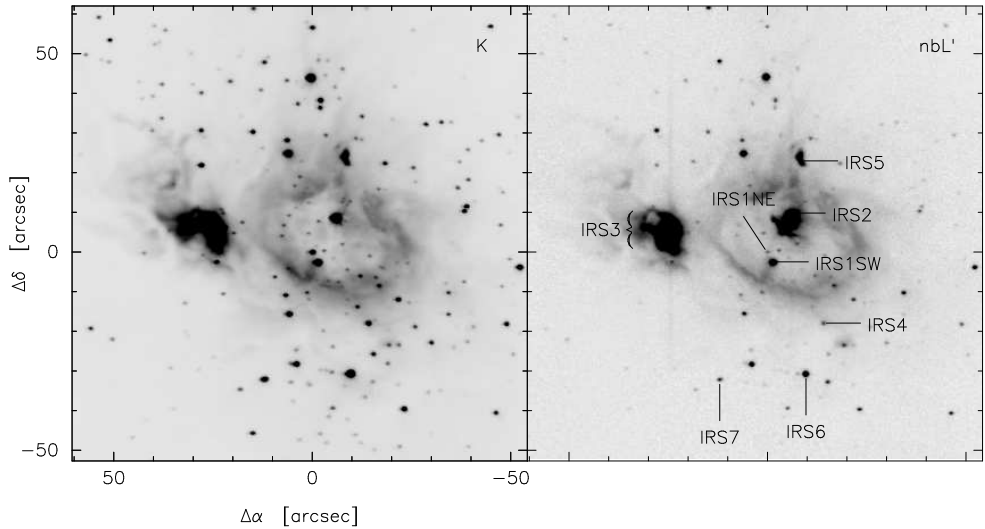}
\end{center}
\caption{
  Location of the IRS sources found by \citet{Beckwith76} identified on the
  $K$-band and $nbL$ images from \citet{Carpenter97}. IRS~1 was
  resolved into two point sources by \citet{Aspin90} and \citet{Howard94},
  where IRS~NE is a foreground star unrelated to the cluster.
  IRS~3 was noted to be extended by \citet{Beckwith76}, and subsequently
  resolved into multiple sources (see Fig.~\ref{fig:irs3}).
  \label{fig:monr2_irs}
}
\end{figure}
Five discrete sources
(IRS 1-5) were detected at 10~\micron, with two additional sources (IRS~6
and IRS~7) appearing at $H$- and $K$-band only. Figure~\ref{fig:monr2_irs}
identifies these sources on modern images of the cluster. IRS~1 has
been resolved into two components \citep{Aspin90,Howard94}, but the fainter of
the two components in the mid-infrared (IRS~1NE) is a foreground field star.
IRS~3 appeared extended in the \citet{Beckwith76} images, and subsequent
observations resolved the source into two point sources
\citep{Howell81,Dyck82,McCarthy82}. \citet{Koresko93} showed that the bright
southern source in IRS~3 is surrounded by a bright conical nebula, which they
suggest is produced by starlight scattered from a circumstellar disk.
\citet{Preibisch02} resolved the IRS~3 region into 6 discrete sources.
\begin{figure}[htb]
\begin{center}
\includegraphics[angle=0,scale=0.75]{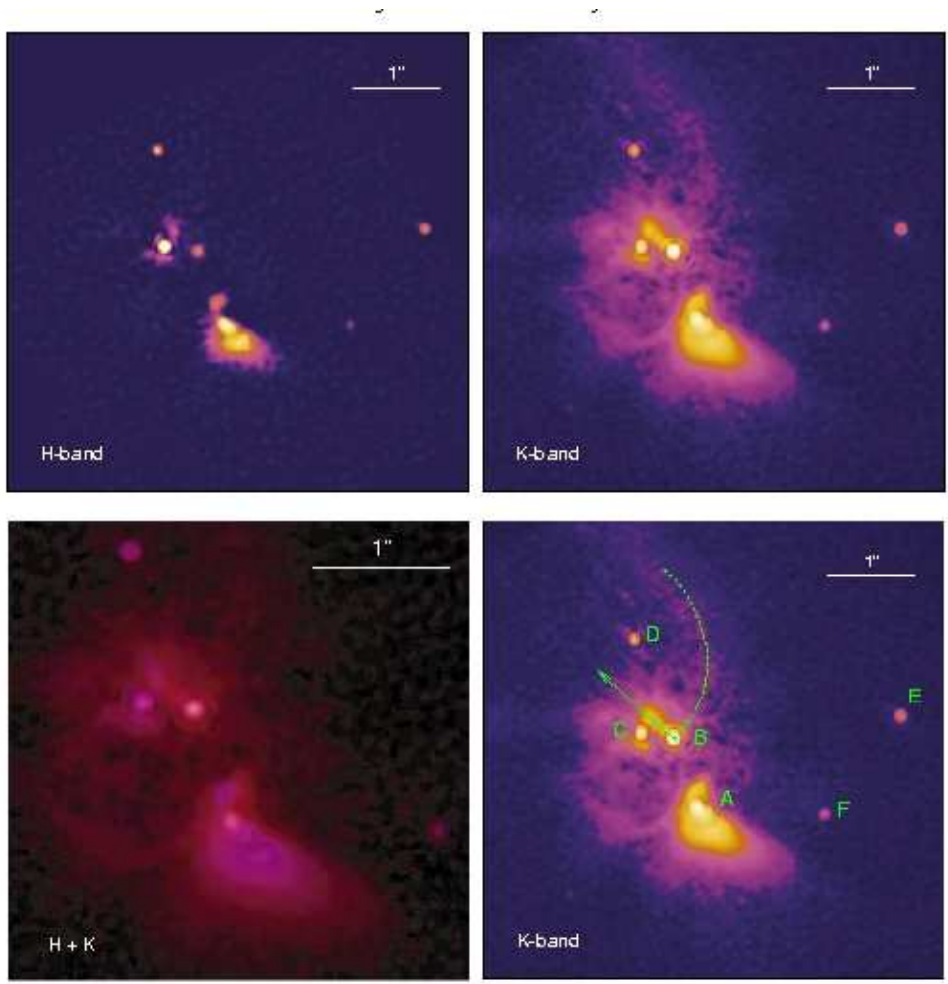}
\end{center}
\caption{
  $H$- and $K$-band images of IRS~3 from speckle interferometry.
  For all images, north is up and east is to the left.
  {\it Upper left}: Color representation of the $H$-band image.
  {\it Upper right}: Color representation of the $K$-band image.
  {\it Lower left}: Color composite of the $H$- and $K$-band images.
  {\it Lower right}: $K$-band image marked with six point sources and
  a possible jet originating from IRS~3B.
  Figure from \citet{Preibisch02}.
  \label{fig:irs3}
}
\end{figure}
Speckle images from their study are presented in Figure~\ref{fig:irs3}.
In addition to the conical nebula around IRS~3A, these images reveal three
knots of emission distributed along a line connecting to IRS~3B.
\citet{Preibisch02} suggest these knots are related to a jet originating
from IRS~3B. The orientation of the possible jet is roughly perpendicular to
the large scale molecular outflow originating from the Mon~R2 core
\citep{Wolf90}, but it may be related to a second outflow that is inferred
from analysis of the gas kinematics \citep{MRL91}. Infrared spectra of IRS~3
at 3.3 and 11.2~\micron\ reveal absorption features attributed to
polycyclic aromatic hydrocarbons \citep{Sellgren95, Bregman00}.

\begin{figure}[ht]
\begin{center}
\includegraphics[angle=0,scale=0.59]{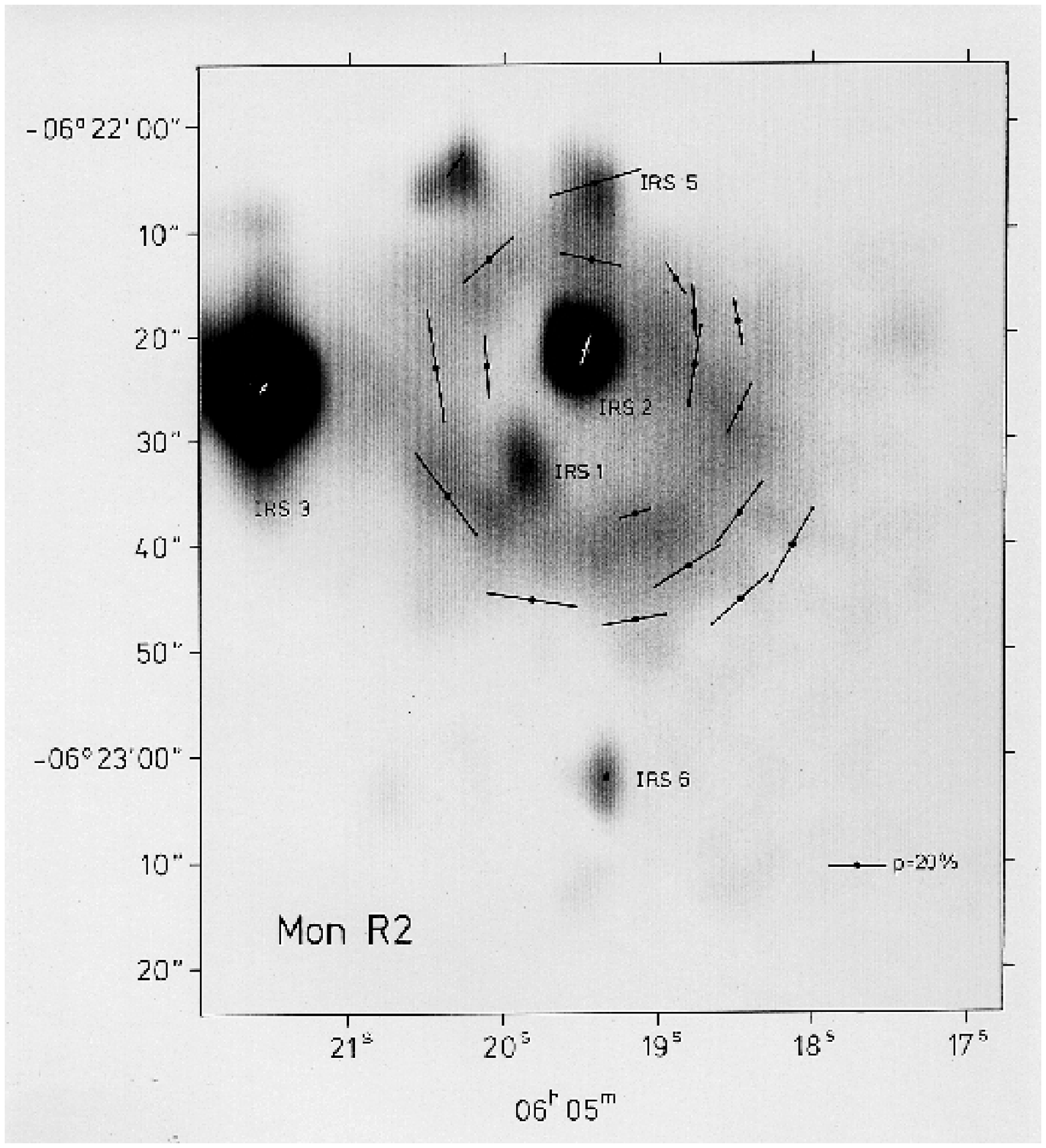}
\end{center}
\caption{
  $K$-band image of the Mon~R2 star forming core superposed with polarization
  vectors measured in $K$-band with 8.4$''$ resolution \citep{Hodapp87}.
  The polarization vectors reveal that the extended emission at 2.2~\micron\
  is a shell-like reflection nebula illuminated by IRS~2.
  \label{fig:monr2_hodapp}
}
\end{figure}

The \citet{Beckwith76} maps revealed a ringlike nebula surrounding the
infrared point source IRS~2 and also including the source IRS~1 (see
Fig.~\ref{fig:monr2_irs}). Linear polarization measurements in the $K$-band
by \citet[][see Fig.~\ref{fig:monr2_hodapp}]{Hodapp87} using a single-beam
infrared polarimeter demonstrated that this nebulosity exhibits strong
polarization in a centro-symmetric pattern centered on IRS~2, identifying it
as a reflection nebula illuminated by IRS~2.
\citet{Aspin90} confirmed these
results using infrared imaging polarimetry with higher spatial resolution. In
addition to the scattered component, they also identified a polarization
component caused by aligned grains that explains the polarization of IRS~2
itself and of the faint extended flux in the reflection nebula close to the
line of sight towards IRS~2.
Additional near-infrared imaging polarimetry was
reported by \citet{Yao97}, who, in addition to confirming the earlier results
on the illumination by IRS~2, also discuss in detail the scattered light
polarization pattern around IRS~3, the brightest of the near-infrared sources.
Their measurements show IRS~3 to have the polarization pattern of an object
with a multi-scattering disk oriented in SE-NW orientation, and larger degrees
of polarization caused by single scattering in the direction perpendicular to
the disk, consistent with the details of the spatial structure of IRS~3
revealed by \citet{Preibisch02}.

The 10~$\mu$m and 20~$\mu$m maps of \citet{Hackwell82} show that, while IRS~2
is the brightest source in the area of the reflection nebula at a wavelength
of 10~$\mu$m, IRS~1 is, by far, the dominant source at 20~$\mu$m, and
appears extended. The VLA observations by \citet{Massi85} revealed a region of
radio-continuum emission coinciding with the near-infrared reflection nebula,
but with a sharp concentration of the radio flux near the southeastern edge
of the region, in the general area of IRS~1.

\begin{figure}[!b]
\begin{center}
  \includegraphics[angle=-90,scale=0.75]{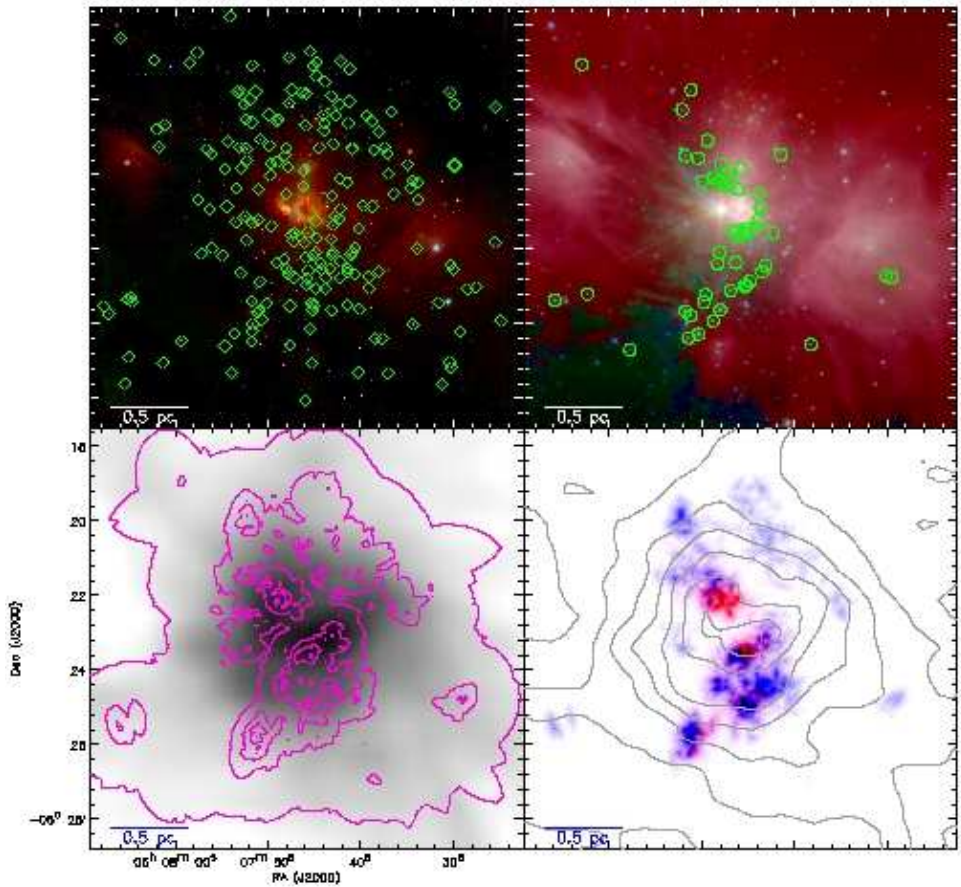}
\end{center}
\caption{
Images of Mon~R2 overlaid with the distribution
of YSOs identified and classified on the basis on $H$, $K$ and Spitzer
3.6 and 4.5~$\mu$m photometry. The upper left is a $JHK$ false color
image with Class II sources marked by green diamonds. The image
in the upper right is the Spitzer/IRAC [3.6][4.5][8.0] false color
image with Class I sources marked by green circles.
The bottom left shows the stellar surface density contours (all YSOs)
overlaid on a greyscale $^{13}$CO image. The bottom right shows
the distribution of Class I and II source color coded in red and blue,
with $^{13}$CO contours overlaid.
Figure from \citet{gut05}.
  \label{fig:monr2_Spitzer}
}
\end{figure}

Far-infrared \citep{Thronson78,Thronson80} and submillimeter \citep{Henning92}
observations have identified IRS~1, IRS~2 and IRS~3 as the most luminous
sources in the cluster. From radiative transfer models, \citet{Henning92}
estimate the luminosities of these three components to be 3000~L$_\odot$,
6500~L$_\odot$, and 14,000~L$_\odot$\ respectively. IRS~1 is considered the ionizing
source of the radio continuum emission; the $K$-band position of IRS~1 is
offset by \about 4\arcsec\ from the peak radio continuum position, which can be
explained in the blister \ion{H}{II} region model \citep{Massi85}. The inferred
spectral type of IRS~1 based on the radio continuum emission is B0 ZAMS, which
is broadly consistent with the infrared luminosity. \citet{Tafalla97}
did not detect radio continuum emission toward IRS~3 even though it is the
most luminous source, and they suggested IRS~3 is in a younger stage of
evolution relative to IRS~1.

Near-infrared images with array detectors have identified a cluster of a few
hundred stars in the Mon~R2 core \citep{Aspin90,Howard94,Hodapp94,Carpenter97}.
Figure~\ref{fig:monr2_color} presents a 3-color composite image of a
$15'\times15'$ region centered on the cluster from \citet{Carpenter97}.
Analysis of the star counts in this image indicates that the cluster extends
over a
$4.5'\times8.5'$ (2.1~pc$\times$2.1~pc) region elongated in a north-south
direction.
Many of the
cluster members have been detected in X-rays with deep {\it Chandra}
observations \citep{Kohno02,Nakajima03}.

Based on analysis of the near-infrared photometric data, \citet{Carpenter97}
found that the stellar mass function in the Mon~R2 cluster is consistent with
the Miller-Scalo mass function for stellar masses $\ge 0.1$ M{$_{\odot}$}. No
compelling evidence for mass segregation was found within the cluster for
stellar masses \aboutless 2 M{$_{\odot}$}, but as has been noted in NGC~2024 and the
Orion Nebula Cluster, the most massive stars are situated near the cluster
center \citep[see][and references therein]{Carpenter97}. \citet{Andersen06}
used HST NICMOS observations to extend the mass function to 20~M$_{\rm Jup}$
by computing the ratio of the number of low mass stars between 0.08 and
1 M{$_{\odot}$} to the number of brown dwarfs between 0.02 and 0.08 M{$_{\odot}$}. Their
results show that this ratio is similar to that inferred in Taurus,
IC~348, the Orion Nebula Cluster, and the system field IMF \citep{Chabrier03}.

Recently, \citet{gut05} studied the distribution of YSOs of different
evolutionary states in a number of embedded young clusters, including Mon~R2.
Figure~\ref{fig:monr2_Spitzer} summarizes these results and compares the
YSO spatial distribution to the distribution of $^{13}$CO emission.
These results demonstrate that the youngest (Class I) YSOs are distributed
in a non-symmetric distribution that strongly resembles the distribution
of the gas. In contrast, the presumed older Class II sources appear more
widely distributed.

\begin{figure}[p]
\begin{center}
\includegraphics[angle=0,width=\textwidth]{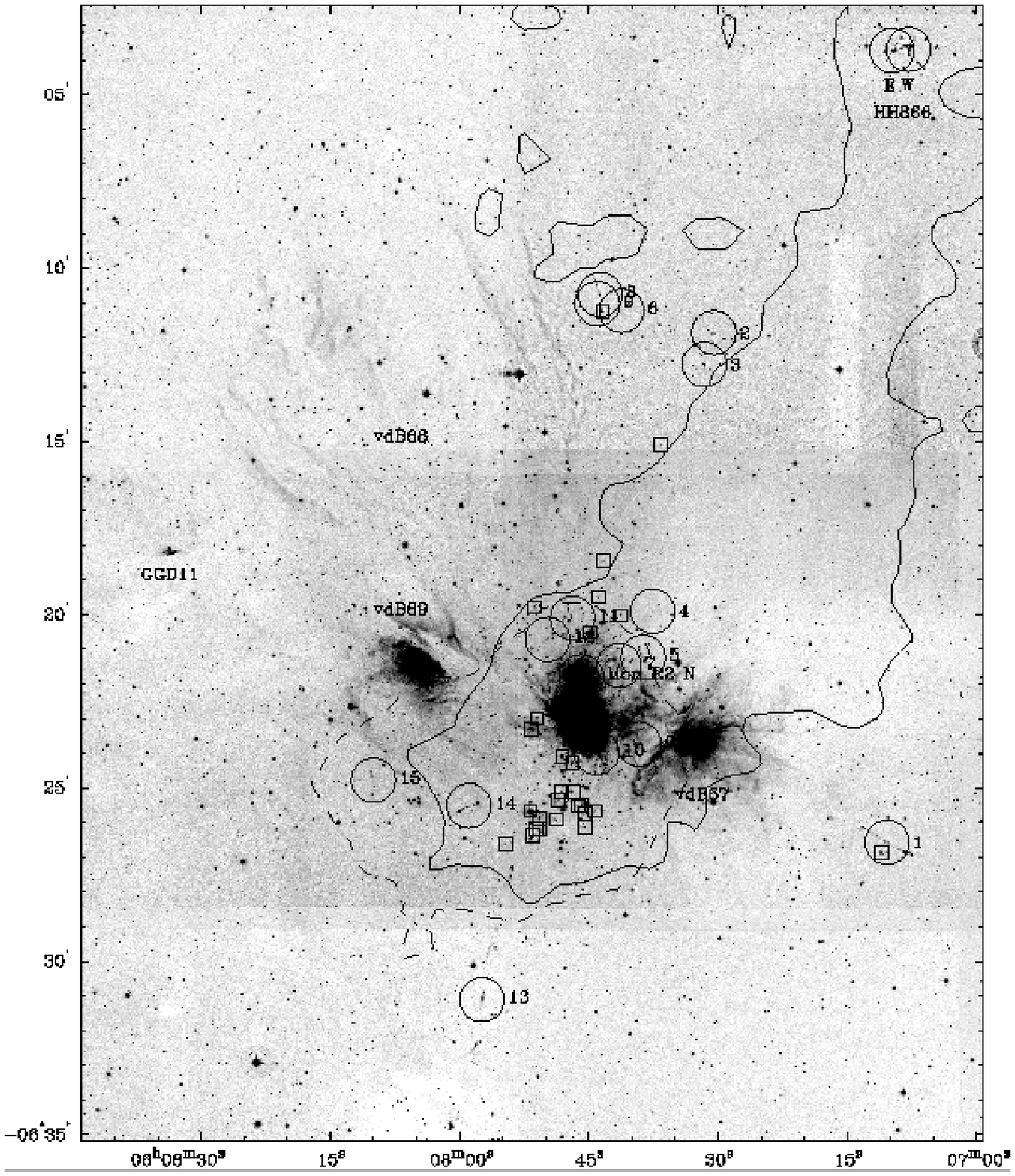}
\end{center}
\caption{
Image of the Mon~R2 star forming region in the 1--0 S(1) emission line of H$_2$ at 2.12~$\mu$m, obtained with WFCAM at UKIRT. The circles
indicate the positions of the newly found H$_2$ jets. The Herbig-Haro object
HH~866 is visible in the north-west corner of the image. The small reflection
nebula near the eastern edge of the
image is GGD~11 \citep{GGD78}. Newly found small reflection nebulae are
indicated by squares. Superposed on this image are lowest contours of the
blueshifted (--2 to 6 km s$^{-1}$, solid line) and redshifted
(14 to 22 km s$^{-1}$, dashed line) CO emission from the map by \citet{Wolf90}.
  \label{fig:jets}
}
\end{figure}
\begin{figure}[p]
\begin{center}
\includegraphics[angle=0,scale=1.0]{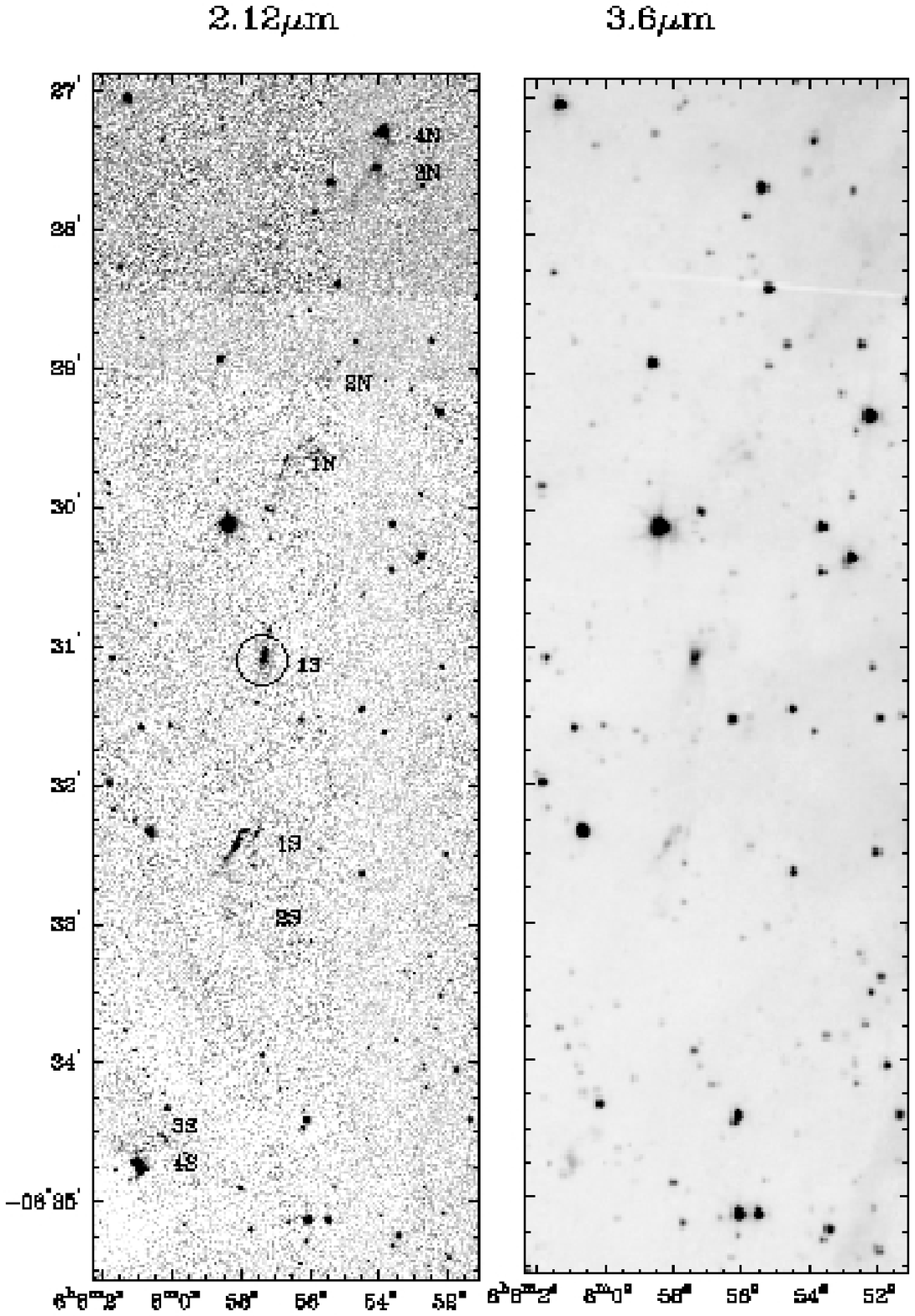}
\end{center}
\caption{
The largest of the 15 H$_2$ jets recently found by \citet{Hodapp07}
in a 2.12~$\mu$m survey of the Mon~R2 region. The left panel shows
the jet HOD07-13 in the H$_2$ 1--0 S(1) line, and the right panel shows
the Spitzer/IRAC 3.6~$\mu$m image.
  \label{fig:HOD07_13}
}
\end{figure}

An imaging survey in the 2.12~$\mu$m emission line of H$_2$ 1--0 S(1) by
\citet{Hodapp07} led to the discovery of 15 new H$_2$ jets in Mon~R2
that represent the very youngest objects in their main accretion phase
(Class 0 and I).
The newly discovered H$_2$ jets are overlaid on
Figure~\ref{fig:jets} as open circles with
labels. Also overlaid are the lowest contours of the blue and redshifted
features in the CO map of \citet{Wolf90}, showing the outline of the
huge outflow that dominates the Mon~R2 cloud. This outflow
is one of the most massive CO outflows known and has carved out a large
cavity in the core of the Mon~R2 molecular cloud.
Only two of the newly found H$_2$ jets are lying clearly outside of the CO
outflow contours.  The longest of the newly
discovered H$_2$ jets (HOD07-13) lies just south of the redshifted CO lobe.
It is shown at 2.12~$\mu$m and 3.6~$\mu$m (Spitzer/IRAC) wavelengths in
Figure~\ref{fig:HOD07_13}. While the Spitzer 4.5~$\mu$m band generally shows
H$_2$ shock-excited emission better than the 3.6~$\mu$m band, those longer
wavelength data were not available when this figure was prepared.

The CO outflows probably have a shell structure with a relatively empty
cavity near the outflow axis \citep{MRL91}. Since the outflowing material
interacts
turbulently with the molecular material of the ambient cloud, triggered
star formation would be expected near the interface between the shell and the
ambient cloud.
The surface density of triggered star formation sites is expected to be
higher in projection along the shell wall compared to the front and back sides
of the outflow shell.
The overall distribution of newly found H$_2$ jets outside of the central
cluster roughly matches this expectation.

\citet{Hodapp07} also found that the area near the Mon~R2 cluster contains
numerous small patches of nebulosity, often with bipolar or cometary
morphology, that did not fit the adopted criteria for identification as shocks
or jets. These are indicated by small square symbols in Fig.~\ref{fig:jets}.
The bipolar or cometary shape of many of these objects suggests that these are
young, embedded stars still surrounded by disks and that they have just
excavated an outflow cavity in the surrounding molecular material that is now
observed via scattered light. These objects appear particularly numerous in the
area about 2 to 3 arcminutes south of the main Mon~R2 cluster, and just north
of the cluster. This finding is consistent with the result by \citet{gut05}
that the Class I sources in Mon~R2, identified by their $J$, $H$, and $K$, and
Spitzer IRAC colors, are concentrated in a filamentary distribution to the
south of the cluster center, and to the north and north-east of the cluster.
While there is no strict relationship between reflection nebulae of bipolar or
cometary morphology and SED Class I, the two criteria cover objects of similar
evolutionary status at the end of their accretion phase.

\begin{figure}[!h]
\begin{center}
  \includegraphics[angle=0,width=0.7\textwidth]{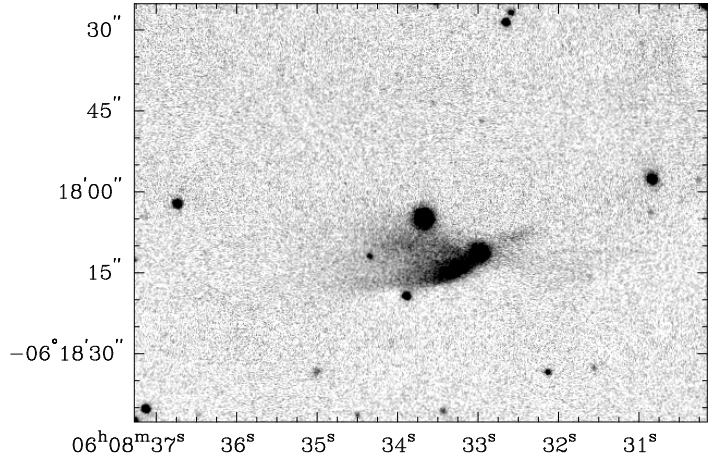}
\end{center}
\caption{
Image at 2.12~$\mu$m of the bipolar reflection nebula GGD~11 from
\citet{Hodapp07}.
  \label{fig:ggd11_hodapp}
}
\end{figure}

To the east of the Mon~R2 cluster, the isolated object GGD~11 has the
morphology of a bipolar reflection nebula, indicating a recently formed young
star in relative isolation from the cluster (see Fig.~\ref{fig:ggd11_hodapp}).

\subsection{GGD 12-15}
\label{ggd12}

\begin{figure}[tb]
\begin{center}
  \includegraphics[angle=-90,scale=0.55]{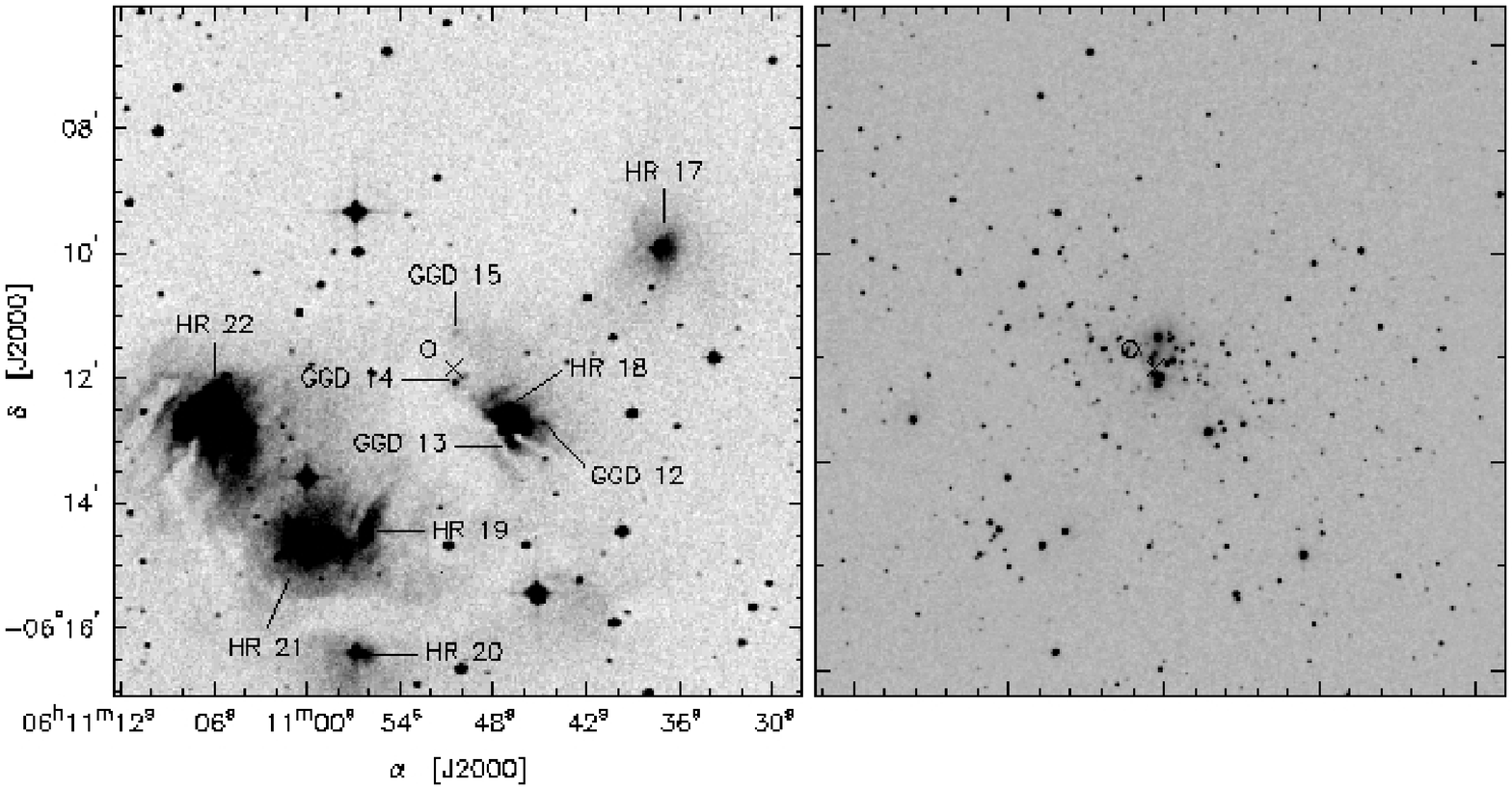}
\end{center}
\caption{
  Images of the GGD 12-15 region. The left panel shows the red print from the
  Palomar Observatory Sky Survey, and the right panel is the 2MASS
  $K_{\rm s}$-band image. The optical nebulae in the GGD 12-15 regions are
  indicated on the Palomar image, where ``HR'' refers to the nebulae listed in
  \citet{Herbst76}, and ``GGD'' to
  \citet[][see also Rodriguez \etal~1980]{GGD78}. In both panels, the cross
  and the open circle indicate the position of a \ion{H}{II} region and water
  maser respectively. \label{fig:ggd12}
}
\end{figure}

The dense core associated with GGD~12-15 possesses similar properties as the
main core in the Mon~R2 cloud. The core extends over a \about 0.7~pc region
as traced by HCO$^+$ and $^{13}$CO J=1-0 with a mass of \about 600-800 M{$_{\odot}$}
\citep{Heaton88}. Submillimeter continuum observations have detected a
compact clump (size \about 0.15~pc) in the center of the core with a mass
of \about 280 M{$_{\odot}$} \citep{Little90}. \citet{Rodriguez82} detected blue and
red shifted CO emission from the core indicative of high velocity gas in
a molecular outflow.

\begin{figure}[htb]
\begin{center}
  \includegraphics[angle=-90,scale=0.75]{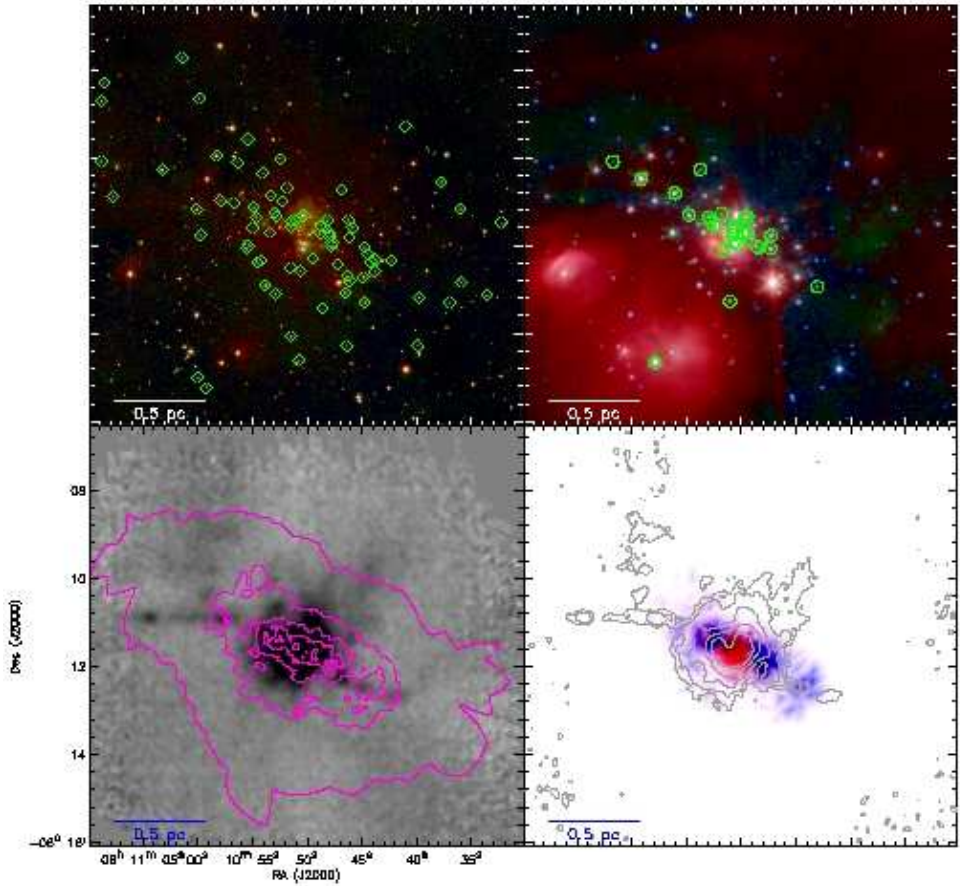}
\end{center}
\caption{
Images of GGD 12-15 overlaid with the distribution
of YSOs identified and classified on the basis of $H$, $K$ and Spitzer
3.6 and 4.5~$\mu$m photometry. The upper left is a $JHK$ false color
image with Class II sources marked by green diamonds. The image
in the upper right is the Spitzer/IRAC [3.6][4.5][8.0] false color
image with Class I sources marked by green circles.
The bottom left shows the stellar surface density contours (all YSOs)
overlaid on a greyscale SCUBA 850~$\mu$m image. The bottom right shows
the distribution of Class I and II sources color coded in red and blue,
with SCUBA 850~$\mu$m contours overlaid.
Figure from \citet{gut05}.
  \label{fig:GGD1215-spitzer}
}
\end{figure}

The embedded cluster in the GGD 12-15 region began to be uncovered first
by \citet{Cohen80}, and, soon after, by \citet{Reipurth83} who found a
group of seven sources \citep[see also][]{Olofsson85}. More recent results
suggest the cluster population is as high as \about 130 stars
\citep{Hodapp94,Carpenter00}.
Figure~\ref{fig:ggd12} presents an optical
and 2MASS $K_{\rm s}$-band image of the GGD~12-15 region. The various optical
nebulae are identified, as well as the position of a compact \ion{H}{II}
region \citep[cross;][]{Rodriguez80,Kurtz94,Tofani95,Gomez98} and
a water maser \citep[open circle;][]{Rodriguez80,Tofani95}.
Recently, \citet{gut05} studied the distribution of YSOs of different
evolutionary state in GGD~12-15. Figure~\ref{fig:GGD1215-spitzer} summarizes
the results and demonstrates that the youngest (Class I) YSOs are distributed
in a non-symmetric pattern that still strongly resembles the distribution
of gas and dust. In contrast, the presumed older Class II sources are
distributed more widely.

The compact \ion{H}{II} region in GGD~12-15 is coincident with mid- to
far-infrared and submillimeter emission \citep{Harvey85,Little90,Persi03}.
The implied bolometric luminosity is $<$6600~L$_\odot$ for a distance of 830~pc,
where the upper limit is implied since multiple sources contribute to the flux
within the \IRAS\ beam \citep{Harvey85,Persi03}. The inferred spectral type
of the ionizing star from the radio continuum flux is B0.5, which is
sufficient to produce most of the observed far-infrared luminosity.
The water maser in GGD~12-15 is offset from the compact \ion{H}{II} region
and is coincident with a compact 20\micron\ source \citep{Harvey85}.
The embedded star associated with the maser is centered between red-
and blue-shifted molecular outflow \citep{Rodriguez82,Little90} lobes and
is the likely source of the outflow \citep{Harvey85}.

\begin{figure}[htbp]
\begin{center}
  \includegraphics[angle=0,scale=0.5]{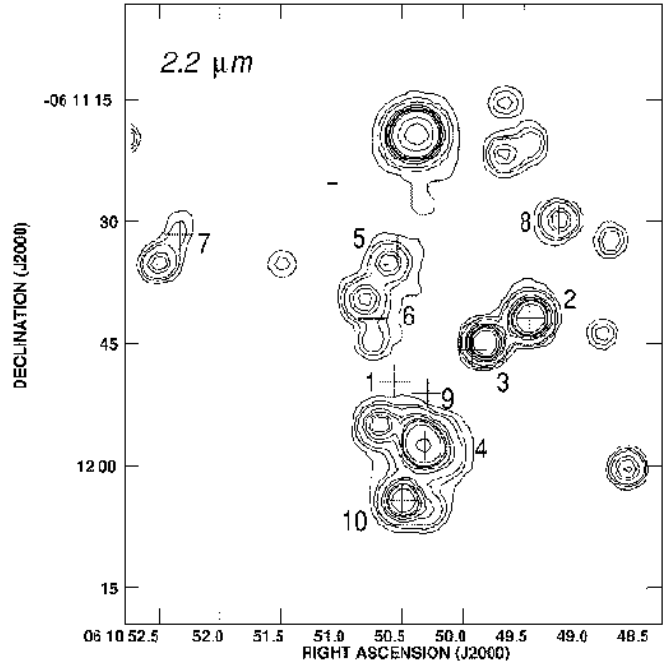}
\end{center}
\caption{
  Contour map of the 2MASS $K_{\rm s}$-band image of the GGD~14 star forming
  region. Crosses mark the position of VLA radio continuum sources from
  \citet{Gomez02}. VLA~1 is a bright cometary \ion{H}{II} region excited by
  an early B-star. Figure from \citet{Gomez02}.
  \label{fig:ggd14_gomez}
}
\end{figure}

\citet{Gomez00} report sensitive VLA radio continuum observations of the
GGD~12-15 region and detected five new radio continuum sources in addition
to the luminous compact \ion{H}{II} region detected in previous surveys.
They originally suggested these fainter sources are ultracompact \ion{H}{II}
regions around B2-B3 stars. More sensitive follow-up VLA observations raised
the total to ten radio continuum sources in the GGD~14 core
\citep[][see Fig~\ref{fig:ggd14_gomez}]{Gomez02}. These additional
observations showed, however, that several of these sources are
time variable, and have negative spectral indices, which led \citet{Gomez02}
to conclude that the radio emission from the fainter sources most likely
originates from gyrosynchrotron radiation from an active magnetosphere.
The two exceptions are VLA~7, which is possibly a radio jet as implied by
the spectral slope of the radio continuum emission \citep{Gomez02}, and
VLA~4, which has a 1-20~\micron\ luminosity of \about 240~L$_\odot$ and may be
B-type star \citep{Persi03}.

\subsection{HH 866 (IRAS 06046-0603)}
\label{HH866}
The Herbig-Haro object HH~866, discovered by \citet{wan05}, is associated
with the \IRAS\ source 06046-0603. The region was
identified by \cite{Carpenter00} as a region of enhanced star density in 2MASS
data, and was identified as a potential CO outflow source by \citet{xu01}.
Images at 2.12~$\mu$m \citep{Hodapp07} shows a system of emission features that can be
morphologically identified as two H$_2$ jets, partly associated with the
optical HH objects.

\begin{figure}[htbp]
\begin{center}
  \includegraphics[angle=-90,scale=0.95]{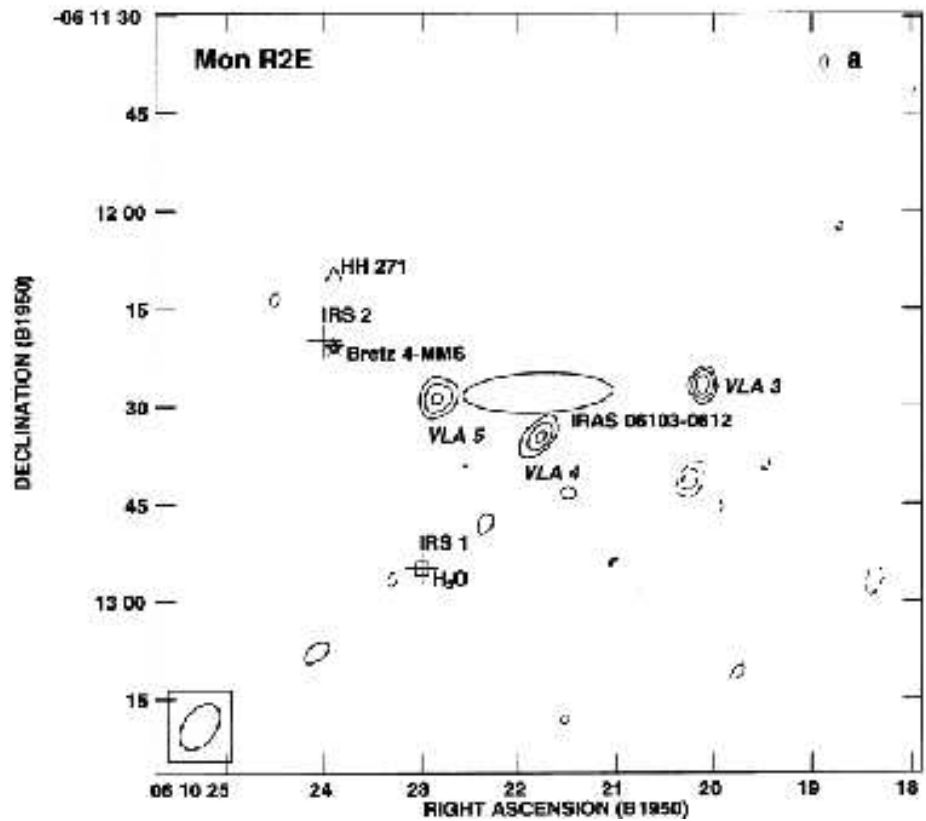}
\end{center}
\caption{
  VLA 6~cm contour map of the GGD~16-17 region (also known as Mon~R2E).
  The ellipse denotes the position uncertainty of the \IRAS\ source.
  The symbols represent the location of the Herbig-Haro object HH~271
  ({\it triangle}), near-infrared sources ({\it plus signs}), $\rm{H_2O}$
  masers ({\it squares}), and the T Tauri star ({\it star}).
  Figure from \citet{Beltran01}.
  \label{fig:beltran}
}
\end{figure}

\subsection{GGD 16-17}
\label{ggd17}
A smaller site of ongoing star formation in the Mon~R2 molecular cloud are the
objects associated with GGD~16 and 17 \citep{GGD78}. To the south of GGD~17,
the T Tauri star Bretz~4 is probably associated with the GGD object.
This star has been studied spectroscopically by \citet{her72} and
was classified by \citet{Carballo92} as a K4 spectral type with a
class 5 \citep{her62} emission spectrum.
The embedded objects associated with GGD~16 and 17 were first studied by
\citet{Reipurth83} and confirmed by \citet{Carballo88}. The infrared
source IRS~2 is positionally coincident with Bretz~4, while the more deeply
embedded IRS~1 has no optical counterpart and lies between the GGD objects.
A detailed optical study by \citet{Carballo92} showed that GGD~17 is
part of a curved jet extending north of the star Bretz~4 and consisting
of HH 271 (also known as GGD~17 HH1) and HH~272 (GGD~17~HH2), and possibly
also HH~273 (GGD~17 HH3). Nebulosity close to the star shows the typical
morphology of scattered light from an outflow cavity wall. The embedded
infrared objects and optical reflection nebulosity in the general GGD~16-17
region is associated with 850~$\mu$m emission \citep{Jenness95}.
\citet{Beltran01} present VLA 6~cm radio continuum images of the GGD~16-17
region. Their radio continuum map, reproduced in Figure~\ref{fig:beltran},
shows the locations of the radio sources with respect to a water maser,
an \IRAS\ source, and HH 271.

\acknowledgements

We would like to thank Bo Reipurth and the referee, Rob Gutermuth, for
valuable comments on this manuscript. We also thank
Russell Croman (http://www.rc-astro.com) for granting permission to reproduce
Figure~\ref{fig:croman}.

\end{document}